\documentclass[%
reprint,
amsmath,amssymb,
aps,
pra,
]{revtex4-2}

\usepackage{graphicx}% Include figure files
\usepackage{dcolumn}% Align table columns on decimal point
\usepackage{multirow}
\usepackage{bm}% bold math
\usepackage{siunitx}

\usepackage[T1]{fontenc}
\usepackage{tikz}

\begin{document}

%\preprint{APS/123-QED}

\title{Arc and Chicane Bunch Compression Schemes for Hard and Soft X-Ray Free Electron Laser Facilities: A Comparison}

\author{Adam Dixon}
\affiliation{MAX IV Laboratory, Lund, Sweden}
\email{adam.dixon.7536@maxiv.lu.se}

\author{Peter Williams}
\affiliation{STFC Daresbury Laboratory, Warrington, United Kingdom}
\affiliation{Cockcroft Institute, Warrington, United Kingdom}

\author{Sara Thorin}
\affiliation{MAX IV Laboratory, Lund, Sweden}

\author{Tessa Charles}
\affiliation{ANSTO-Australian Synchrotron, Melbourne, Australia}

\author{Alexander Brynes}
\affiliation{STFC Daresbury Laboratory, Warrington, United Kingdom}
\affiliation{Cockcroft Institute, Warrington, United Kingdom}

\author{Ian Bailey}
\affiliation{Lancaster University, Lancaster, United Kingdom}
\affiliation{Cockcroft Institute, Warrington, United Kingdom}

\author{Andrzej Wolski}
\affiliation{University of Liverpool, Liverpool, United Kingdom}
\affiliation{Cockcroft Institute, Warrington, United Kingdom}

\date{\today}% It is always \today, today,

\begin{abstract}
	
	X-ray free-electron laser (XFEL) facilities require progressive compression of electron bunches as they are accelerated from an injector to the undulators. This is necessary to achieve the peak currents required for efficient lasing, without compromising transverse brightness. In the present generation of XFELs, high peak currents are achieved by means of a sequence of four-dipole bunch compression chicanes. It is well known that these systems are not ideal in that they allow projected emittance dilution at the percent level, and they exhibit amplification of microbunching, which typically must be controlled through the otherwise unwanted addition of slice energy spread by use of a laser heater. Both emittance dilution and microbunching are mediated through coherent synchrotron radiation that occurs within a bunch compression chicane. In this paper we introduce a new option for bunch compressors, that of full arc compression, and compare it to the standard four-dipole chicane and to a recently proposed variant, the five-dipole CSR mitigating chicane. It is shown that the arc compressor and the five-dipole chicane are able to give greatly improved XFEL performance compared to the standard four-dipole chicane, both in soft and hard X-ray regimes. This is demonstrated in the context of two proposed XFELs, SXL at MAX-IV, Sweden, and UK-XFEL. It is further shown that the optimal choice of compression option depends on the particular FEL scheme. This means that a simultaneous multi-FEL facility, such as UK-XFEL, must implement both arc and five-dipole methods and must be able to select between them on a bunch-by-bunch basis. %Finally, a beamline system is proposed to achieve this.
	
\end{abstract}

\maketitle

\section{Introduction}
Free electron lasers (FELs) are sources of extremely bright photon beams that are used to study small-scale structures and ultra-fast dynamical processes \cite{freundPrinciplesFreeElectron2018,pellegriniPhysicsXRayFreeElectron2016,SoftXrayLaser2021,UKXFELScienceCase2020,corkumAttosecondScience2007,mcneilXRayFreeElectronLasers2010}. The photon beam can be characterised by the spectral brightness, which strongly depends on the properties of the electron bunches used to generate the FEL pulses, such as the bunch charge, bunch length, transverse emittance, and energy spread.  To ensure efficient lasing, a high peak current is needed, which must be achieved using bunch compressors.  However, the transverse emittance and energy spread of the electron beam are then often limited by collective effects in the bunch compressors; a particularly important collective effect in modern FELs is coherent synchrotron radiation (CSR) \cite{saldinAnalyticalDescriptionLongitudinal2002,dowellCoherentSynchrotronRadiation1998}, which can drive emittance growth and microbunching instability (MBI) \cite{saldinLongitudinalSpaceChargeDriven2004,huangEffectsLinacWakefield2003,heifetsCoherentSynchrotronRadiation2002}. The impact of CSR and other collective effects have been studied at multiple facilities \cite{baneMeasurementsModellingCoherent2009,dimitriSuppressionMicrobunchingInstability2010,dimitriMicrobunchingInstabilityStudy2017,brynesLimits1DCoherent2018,brynesAddendumLimits1D2021,pratEnergySpreadBlowup2022}. Developments in bunch compressor geometries and optics have led to designs that have been shown to be effective at mitigating emittance growth due to CSR and energy spread blow-up due to MBI \cite{khanNovelBunchCompressor2022,zengSelfCancelationCoherentSynchrotron2024,zengSuppressingCoherentSynchrotron2024,douglasSuppressionEnhancementCSRDriven1998,dimitriCancellationCoherentSynchrotron2013,brynesMitigationMicrobunchingInstability2024}. 
Despite the effectiveness of the developments in bunch compressor design it is not necessarily the case that four-dipole/symmetric C-chicane bunch compressors, which are used at a number of successful FEL facilities \cite{arthurLinacCoherentLight2002,altarelliXFELEuropeanXray2006,FERMIElettraConceptualDesign2007,togawaElectronbunchCompressionUsing2009,SwissFELConceptualDesign2011,pratCompactCostEffectiveHard2020,emmaFirstLasingOperation2010,ishikawaCompactXrayFreeelectron2012,deckingMHzrepetitionrateHardXray2020}, are the optimal choice for producing low emittance and high current bunches demanded by future FELs.

In this paper, four-dipole/symmetric C-chicane bunch compressors are compared to two alternative CSR-mitigating configurations, looking in particular at the ability of the different designs to deliver low emittance and high peak current electron bunches to an FEL. First, the three different bunch compressor configurations are discussed. Then, the bunch compressors are modelled in the context of soft and hard X-ray FEL facilities. For each compressor, the electron bunch and FEL pulse properties for various SASE FEL schemes will be compared. The sensitivity of electron bunch properties to variations in charge and compression factor will be evaluated for each compression scheme. Finally, the MBI gains in different schemes will be compared.

\section{Bunch Compressor Designs for CSR Mitigation}
\label{sec:bc_comp_types}

The three bunch compressors that are considered in this paper are: symmetric C-chicane, five-dipole chicane, and arc with optics balance.  Schematics of each type of bunch compressor are shown in Fig.~\ref{fig:all_layouts}.  

\subsection{Symmetric C-chicane}

Symmetric C-chicane bunch compressors use a simple magnetic layout (requiring at least three dipole magnets) to generate longitudinal dispersion, i.e.~a path length variation with energy, which is necessary for bunch compression. While only three dipoles are required, it is more common to use four dipoles (in which the middle dipole of a three-dipole chicane is split into two dipoles and separated by a drift) so that diagnostics or collimators can be included \cite{dohlusBeamDynamicsNewsletter2005}. The design benefits from only needing a common power supply. Variation of the longitudinal dispersion ($R_{56}$), which is necessary for adjusting the compression factor, can be achieved by changing the dipole field strengths while simultaneously adjusting the transverse positions of the central dipoles \cite{borlandHighlyFlexibleBunch2000,williamsDevelopmentsCLARAAccelerator2015}. %A generic four-dipole/symmetric C-chicane is shown in Figure~\ref{fig:all_layouts}. 

%\begin{figure}[ht]
%	%\centering
%	\includegraphics[width=0.8\linewidth]{Figures/sc_chicane.pdf}
%	\caption{Schematic of a symmetric C-chicane bunch compressor.}
%	\label{fig:sc_layout}
%\end{figure}

Emittance growth due to CSR is partially mitigated in a symmetric C-chicane by minimising the optical beta function in the bend plane in the final dipoles of the bunch compressor, where the bunch becomes short, which reduces the contribution of CSR-induced angular spread to the projected emittance \cite{dohlusImpactOpticsCSRRelated2005,dohlusBeamDynamicsNewsletter2005}. This strategy is progressively less effective at strong compression factors, as shown in Ref.~\cite{sahaiPetaVoltsMeterPlasmonics2023}, where the transverse emittance is quickly degraded for large peak currents in a symmetric C-chicane bunch compression scheme for a plasmonic-based accelerator.

\subsection{Five-dipole Chicane}

Five-dipole \cite{khanNovelBunchCompressor2022} or asymmetric S-chicane \cite{zengSelfCancelationCoherentSynchrotron2024} bunch compressors mitigate CSR-induced emittance growth using a modified chicane geometry, in which the last but one dipole (fourth dipole of a five-dipole chicane or third dipole of a four-dipole asymmetric S-chicane) has the opposite bend angle and the dispersion function changes sign compared to the previous dipoles. The opposite bend angle and opposite sign of dispersion results in partial cancellation of the CSR kicks from the preceding dipoles and a reduction in CSR-induced emittance growth.
The optimal geometry for CSR mitigation depends on the first-order longitudinal dispersion, compression factor and length constraints \cite{khanNovelBunchCompressor2022,zengSelfCancelationCoherentSynchrotron2024}. %Figure~\ref{fig:all_layouts} shows a schematic of a five-dipole chicane. 

%\begin{figure}[ht]
%	%\centering
%	\includegraphics[width=0.8\linewidth]{Figures/c5_chicane_annotated.pdf}
%	\caption{Schematic of a five-dipole chicane bunch compressor. Red arrows show the direction and magnitude (not to scale) of the CSR-induced change in horizontal position $\Delta x$ at the second, third and fourth dipole. The CSR kicks would approximately cancel each other.}
%	\label{fig:c5_layout}
%\end{figure}

Five-dipole chicanes have chicane-like first-order longitudinal dispersion and an electron bunch must be accelerated on the rising side of the RF waveform to be compressed. The sign of $R_{56}$ and chirp is the same as in a symmetric C-chicane compression scheme, so the longitudinal dynamics is similar in each of these two types of bunch compressor. Additionally, $R_{56}$ variability can be achieved in the same way as in a four-dipole chicane, although changing the value of $R_{56}$ can impact the effectiveness of CSR mitigation. 

\subsection{Arc Compressors with Optics Balance}

Arc-like bunch compressors, such as those used in the MAX IV linac \cite{thorinExperienceInitialMeasurements2018,SoftXrayLaser2021}, are another alternative to symmetric C-chicanes. Arc-like bunch compressors are distinct from chicane-like compression schemes as the first-order longitudinal dispersion has the opposite sign: this means that electron bunches must be accelerated on the falling side of the RF waveform in order to generate a positive chirp for compression. The positive chirp can also be generated or enhanced by short range wakefields and CSR, which means that bunches can be accelerated closer to the crest of the RF, making more efficient use of the linac \cite{karlbergMAXIVLinac2013}.

Arcs require quadrupoles to control the transverse dispersion, which gives them a unique advantage in that the phase advance is tunable and can be optimised for CSR cancellation: this is referred to as optics balance \cite{douglasSuppressionEnhancementCSRDriven1998,dimitriCancellationCoherentSynchrotron2013,dimitriTransverseEmittancepreservingArc2015,jiaoGenericConditionsSuppressing2014}. %Figure~\ref{fig:all_layouts} shows a schematic of an arc, specifically a double bend achromat, with optics balance.

%\begin{figure}[ht]
%	%\centering
%	\includegraphics[width=0.8\linewidth]{Figures/optics_balance.pdf}
%	\caption{Schematic of a double-bend achromat with optics balance. Red arrows show the direction and magnitude (not to scale) of the CSR-induced change in horizontal position $\Delta x$ at the first and second dipole. The CSR kicks would approximately cancel each other due to the phase advance between the centre of each dipole.}
%	\label{fig:optics_balance_layout}
%\end{figure}

In addition to CSR mitigation, arc bunch compression has other advantages: magnetic linearisation of the longitudinal phase space distribution of particles in a bunch (using harmonic RF) is not needed \cite{thorinExperienceInitialMeasurements2018}; arrival time jitter can be reduced as a result of reduced energy jitter (due to cancellation of phase and amplitude jitter at the `magic angle', which only occurs on the falling side of the RF waveform) \cite{manstenCancellationKlystroninducedEnergy2024}; and in a variable compressor it is possible to achieve both positive and negative values of longitudinal dispersion \cite{williamsArclikeVariableBunch2020,dixonReductionArrivalTime2024}.%; high peak current and short electron bunches easily achieved due to single spike current profiles. 

Potential disadvantages of arc bunch compressors include the fact that the sign of the chirp (correlation between longitudinal position and energy) means the chirp cannot be removed using conventional corrugated dechirpers \cite{baneCorrugatedPipeBeam2012}.  There are also stronger chromatic effects due to focusing elements, and increased orbit jitter. The latter can be mitigated using higher-order lattice design in combination with modern control systems incorporating feedback. 

%A benefit of magnetic linearisation, which can be applied to both arc-like and chicane-like bunch compressors \cite{thorinExperienceInitialMeasurements2018,charlesCurrentHornSuppressionReduced2017,sudarNonLinearCompressionOctupoles2020}, is that closely separated bunches (within the same RF bucket) can be compressed and linearised in the bunch compressor. This can be useful for two-pulse FEL schemes \cite{SoftXrayLaser2021} or for a driver and witness beam in a plasma wakefield accelerator. Closely separated bunches that must move through a harmonic cavity will experience different voltages, and the longitudinal phase space distribution may not be linearised for both bunches.

\begin{figure*}[ht]
	%\centering
	\includegraphics[width=0.95\linewidth]{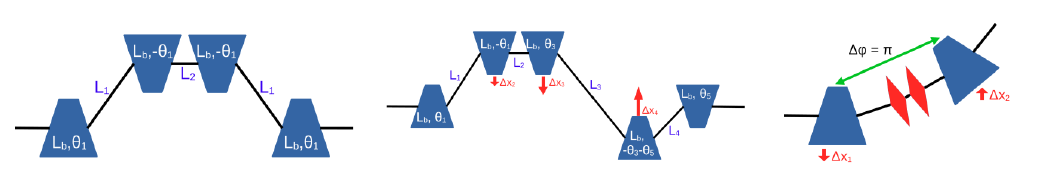}
	\caption{Configurations of three different bunch compressor schemes. Left: symmetric C-chicane. Middle: five-dipole chicane. Right: arc (achromat) with optics balance. Red arrows show the direction and magnitude (not to scale) of the CSR-induced changes in horizontal momentum of particles. The CSR kicks would approximately cancel each other in the five-dipole chicane and double-bend achromat.}
	\label{fig:all_layouts}
\end{figure*}

\section{Application to 3 GeV Soft X-ray FEL: SXL@MAX-IV}
\label{sec:bc_comp_maxiv}

In this section, the three bunch compressors discussed in section~\ref{sec:bc_comp_types} are compared in the context of a normal-conducting S-band linac used to drive a soft X-ray ($0.25 - 1.25 \ \si{\kilo\electronvolt}$) FEL. The design is based on the MAX IV linac and its proposed Soft X-ray Laser \cite{SoftXrayLaser2021}.

The MAX IV linac uses $3\,\si{\giga\hertz}$ normal-conducting RF cavities to accelerate electron bunches to $3\,\si{\giga\electronvolt}$. Electron bunches are compressed in a two-stage compression scheme, first at $265\,\si{\mega\electronvolt}$ and finally at $3\,\si{\giga\electronvolt}$.

\begin{figure}[ht]
	\includegraphics[width=0.99\linewidth]{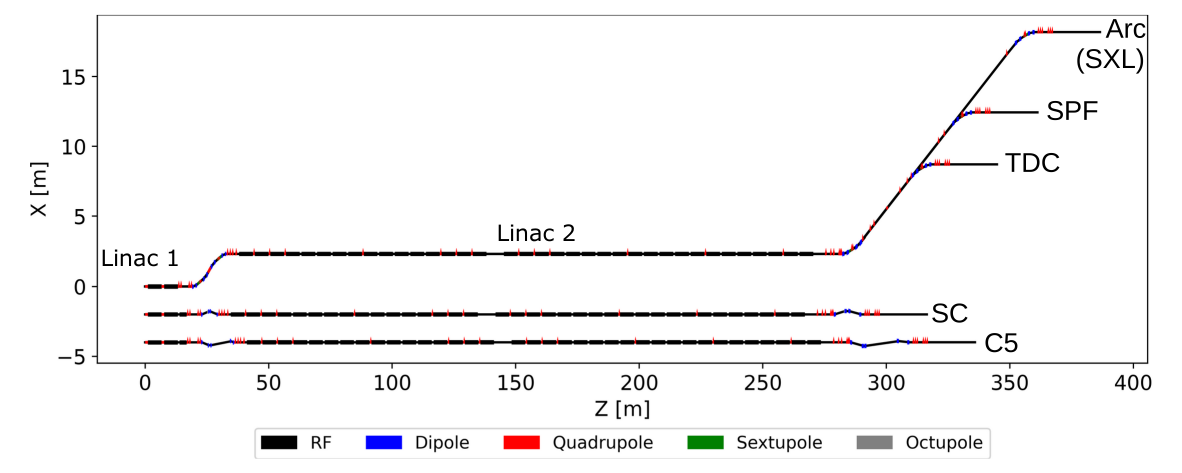}
	\caption{Layout of linac with arc (top), symmetric C-chicane (middle), and five-dipole chicane (bottom) bunch compressors. The arc bunch compression scheme uses the proposed SXL beamline at MAX IV. Existing Short Pulse Facility (SPF) and diagnostics (TDC) beamlines at MAX IV are also shown.}
	\label{fig:sxl_comp_layout}
\end{figure}

The designs of the arc bunch compressors used for the comparison in this section are detailed in the conceptual design report for SXL \cite{SoftXrayLaser2021} and Ref.~\cite{dixonBeamDynamicsBunch2025}. The bunch compressors implement optics balance for CSR mitigation, multiple sextupole pairs for longitudinal phase space linearisation and reduction of higher-order dispersion \cite{bjorklundsvenssonThirdOrderDoubleAchromatBunch2019}, and additional dipoles (not shown in Fig.~\ref{fig:sxl_comp_layout}) for $R_{56}$ variability.

The two chicane bunch compression schemes use bunch compressors which have the same magnitude and opposite sign first-order longitudinal dispersion, $R_{56}$. Additionally, the dipole lengths, and where possible the bend angles, are the same. The geometry of the five-dipole chicane, specifically the parameter space ($L_1$, $L_3$, $L_4$, $\theta_3$ and $\theta_4$ for dipole lengths and bending angles), has been optimised in \texttt{ELEGANT} simulations \cite{borlandElegantFlexibleSDDSCompliant2000} for reduced CSR-induced emittance growth. Linearisation of the longitudinal phase space distribution is achieved with $3^\text{rd}$-harmonic cavities preceding the first bunch compressor.

The layouts of the three accelerators are shown in Fig.~\ref{fig:sxl_comp_layout}. The arc compression scheme results in an $18\,\si{\metre}$ horizontal translation of the beamline at the exit of the second bunch compressor, whereas the two symmetric chicanes result in zero horizontal translation. The large horizontal translation of the second arc bunch compressor allows space for two additional beamlines, one for the Short Pulse Facility (SPF) and another for diagnostics (TDC).

Chromatic effects were investigated for each of the bunch compressors in Ref.~\cite{dixonBeamDynamicsBunch2025}. The results show that the sextupoles and octupoles used in the arc bunch compressors to reduce transverse chromatic effects typically produce stronger coupling between transverse coordinates and energy than either the symmetric C-chicanes or five-dipole chicanes. This leads to variations in the centroids and Twiss parameters in transverse slices along the length of the bunch. However, this has negligible impact on the projected emittance growth of the arc scheme, shown in Tables~\ref{tab:sxl_100pc_bunch_properties} and \ref{tab:sxl_10pc_bunch_properties}.

The impact of CSR on the electron bunch properties, and then on the FEL output, is investigated for two different bunch charges. The two cases use the same bunch charges as modes 1A and 1B proposed in the SXL conceptual design report \cite{SoftXrayLaser2021}, which are $100\,\si{\pico\coulomb}$ and $10\,\si{\pico\coulomb}$, respectively. The electron distributions were produced from \texttt{ASTRA} \cite{floettmannAstraSpaceCharge2017} simulations of the MAX IV injector for SXL. The bunch charges and full-width at ten percent maximum (FWTM) bunch lengths of the two cases studied in this section are:

\begin{enumerate}
	\item[A.] $100\,\si{\pico\coulomb}$ bunch compressed to a final bunch length of $100\,\si{\femto\second}$.
	\item[B.] $10\,\si{\pico\coulomb}$ bunch compressed to a final bunch length of $5\,\si{\femto\second}$.
\end{enumerate}

Details of the compression ratios and RF phase settings used in each compression scheme can be found in Ref.~\cite{dixonBeamDynamicsBunch2025}.

The electron bunch properties achieved in each compression scheme using a bunch charge of $100\,\si{\pico\coulomb}$ are shown in Table~\ref{tab:sxl_100pc_bunch_properties}. An additional arc compression case (arc-2), which has a similar peak current to the electron bunches from the chicane compression schemes, has been included in the comparison. The bunch properties for the $10\,\si{\pico\coulomb}$ bunch charge case are shown in Table~\ref{tab:sxl_10pc_bunch_properties}. Bunch lengths are defined as the full-width at ten percent maximum current.

\begin{table}[t]
	\caption{Electron bunch properties at the entrance of the FEL for each compression scheme using a bunch charge of $100\,\si{\pico\coulomb}$. Two arc compression cases are shown, one has similar bunch length (Arc) and the other has similar peak current (arc-2), compared with the two chicane compression schemes.}
	\begin{ruledtabular}
		\begin{tabular}{l c c c c}
			Bunch properties & Arc & SC & C5 & Arc-2 \\
			\hline
			Energy [\si{\giga\electronvolt}] & 3.0 & 3.0 & 3.0 & 3.0 \\
			Peak current [\si{\kilo\ampere}] & 2.41 & 1.45 & 1.30 & 1.27 \\
			Bunch length [\si{\femto\second}] & 99.95 & 101.71 & 99.69 & 170.58 \\
			$\varepsilon_{n,x/y}$ [\si{\milli\meter \milli\radian}] & 0.42/0.37 & 0.88/0.37 & 0.38/0.37 & 0.42/0.37 \\
			Energy spread [\si{\percent}] & 0.21 & 0.18 & 0.19 & 0.19 \\
		\end{tabular}
	\end{ruledtabular}
	
	\label{tab:sxl_100pc_bunch_properties}
\end{table}

\begin{table}[t]
	\caption{Electron bunch properties at the entrance of the FEL for each compression scheme with bunch charge $10\,\si{\pico\coulomb}$.}
	\begin{ruledtabular}
		\begin{tabular}{l c c c}
			Bunch properties & Arc & SC & C5 \\
			\hline
			Energy [\si{\giga\electronvolt}] & 3.00 & 3.00 & 3.00 \\
			Peak current [\si{\kilo\ampere}] & 3.73 & 3.15 & 3.43 \\
			Bunch length [\si{\femto\second}] & 5.20 & 4.70 & 4.46 \\
			$\varepsilon_{n,x/y}$ [\si{\milli\meter \milli\radian}] & 0.29/0.17 & 0.44/0.17 & 0.33/0.17 \\
			Energy spread [\si{\percent}] & 0.14 & 0.12 & 0.12 \\
		\end{tabular}
	\end{ruledtabular}
	\label{tab:sxl_10pc_bunch_properties}
\end{table}

\begin{figure}[t]
	\includegraphics[width=0.4825\linewidth]{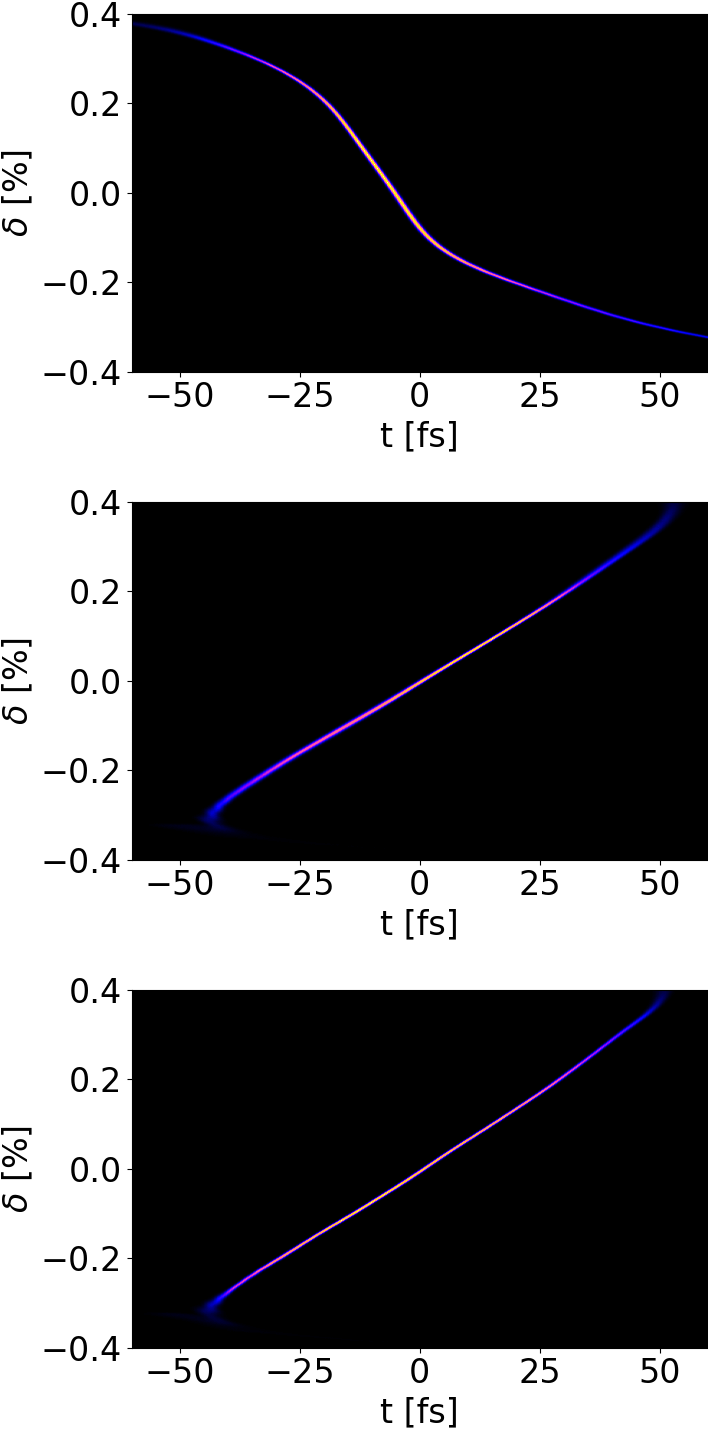}
	\includegraphics[width=0.4975\linewidth]{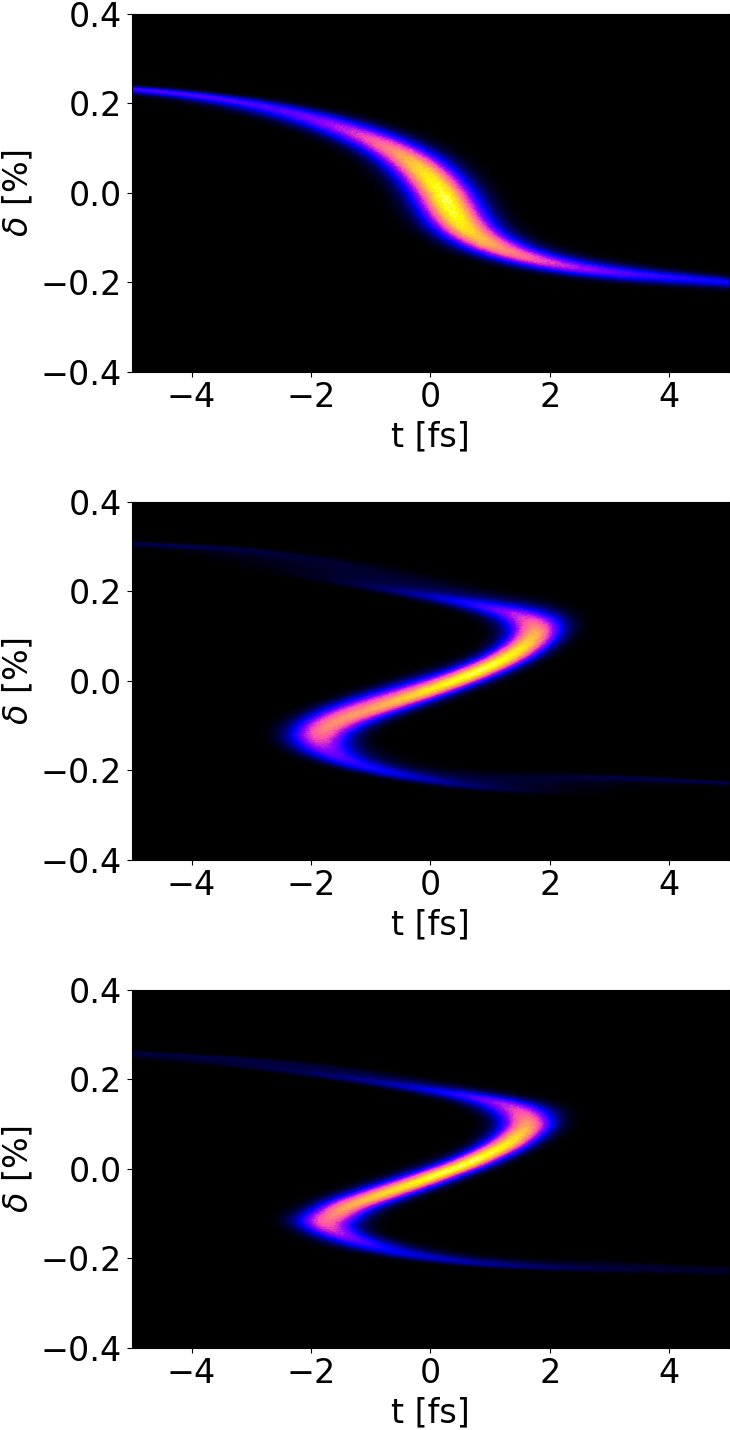}
	\caption{Longitudinal phase space distributions of electron bunches from arc (top), symmetric C-chicane (middle), and five-dipole chicane (bottom) compression schemes for bunch charges: $100\,\si{\pico\coulomb}$ (left) and  $10\,\si{\pico\coulomb}$ (right).}
	\label{fig:sxl_ps}
\end{figure}

\begin{figure}[t]
	\includegraphics[width=0.4825\linewidth]{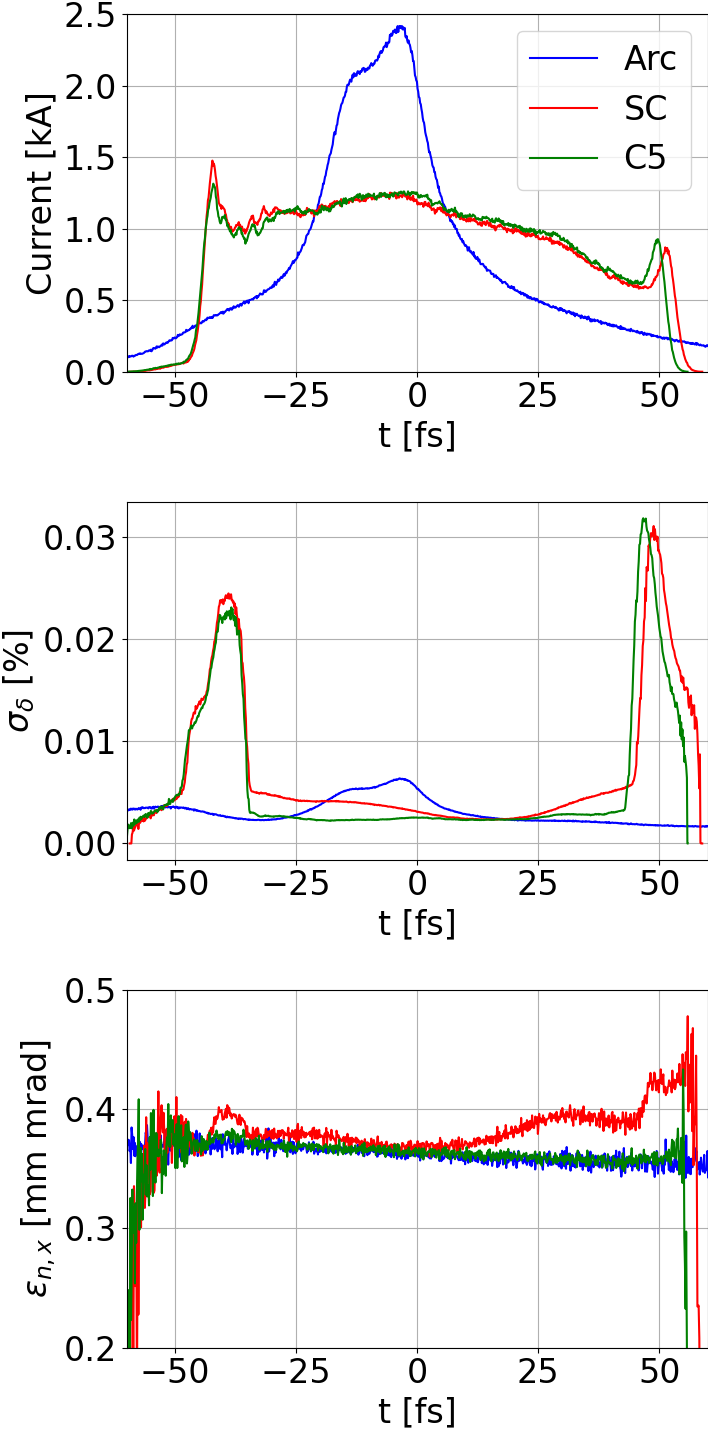}
	\includegraphics[width=0.4975\linewidth]{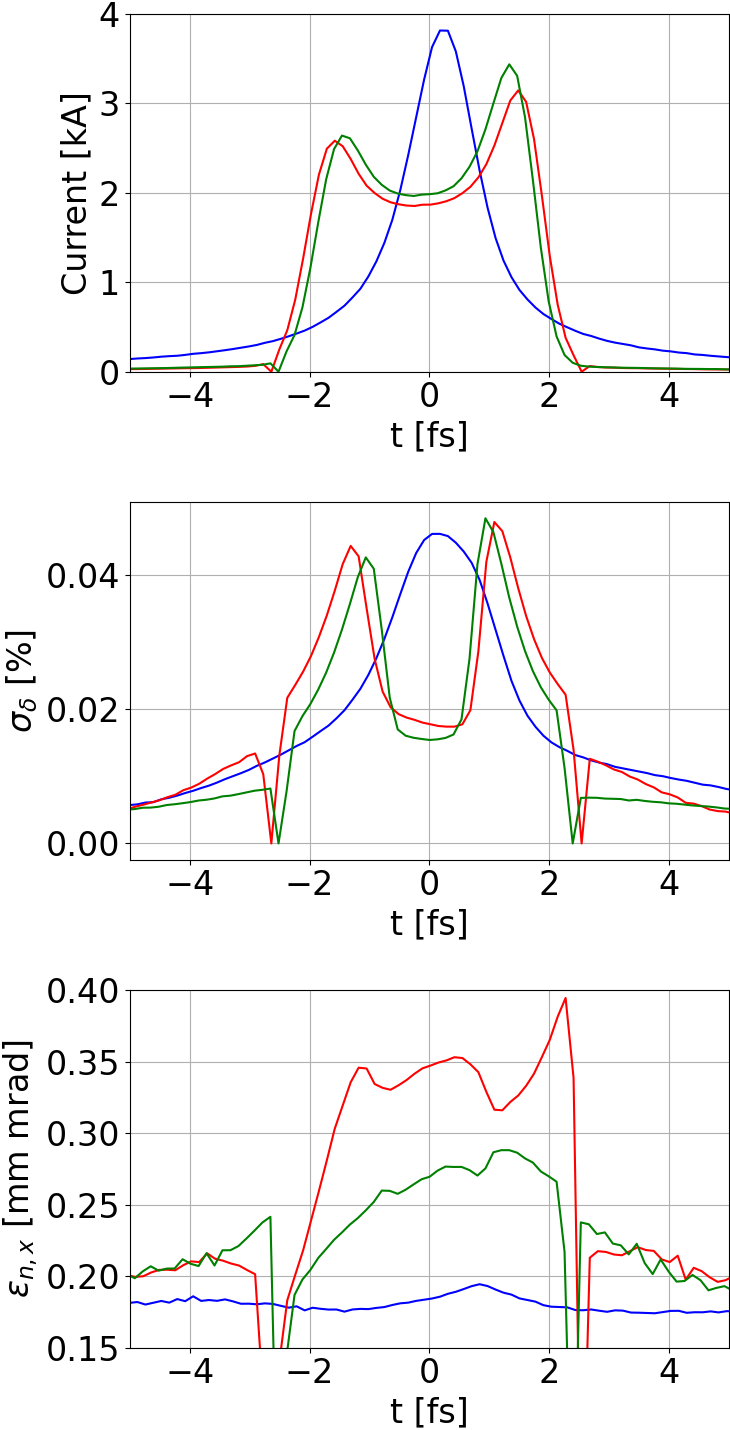}
	\caption{Current (top), slice energy spread (middle) and horizontal slice emittance (bottom) of electron bunches from arc (blue), symmetric C-chicane (red), and five-dipole chicane (green) compression schemes for bunch charges: $100\,\si{\pico\coulomb}$ (left) and  $10\,\si{\pico\coulomb}$ (right).}
	\label{fig:sxl_slice}
\end{figure}

Short range wakefields and CSR enhance the chirp at the core of electron bunches in arc compression schemes (see Fig.~\ref{fig:sxl_ps}) resulting in a stronger local compression factor which produces a single spike in the current profile (see blue curve in Fig.~\ref{fig:sxl_slice}). In the $100\,\si{\pico\coulomb}$ bunch charge case, this single spike in the current profile has a significantly higher peak current than the other compression schemes. However, for the $10\,\si{\pico\coulomb}$ bunch charge case, the variation in peak current between the different compression schemes is much smaller due to the fact that current spikes develop due to folds in the longitudinal phase space distributions of electron bunches from chicane compression schemes. The formation of the folds in the longitudinal phase space distributions can be described with a caustic theory \cite{charlesCausticBasedApproachUnderstanding2016,charlesCurrentHornSuppressionReduced2017}. 

The projected horizontal emittances of electron bunches in arc and five-dipole chicane compression schemes are better preserved than in symmetric C-chicane compression schemes because of the effective mitigation of emittance growth due to CSR. The smaller emittance growth of the $10\,\si{\pico\coulomb}$ electron bunch from the arc scheme compared to the five-dipole chicane scheme could be a result of the single spike in the current profile compared to the double current spike (see Fig.~\ref{fig:sxl_slice}). The single spike at the centre of the bunch in the arc compressor radiates over the head of the bunch and spoils the emittance of the low-current slices; however, electron bunches from the chicane compression schemes have current spikes at the head and tail of the bunches, so the current spike at the tail radiates over the core of the bunch, degrading the emittance of high-current slices. %The symmetric C-chicane compression does not preserve emittance as well as either five-dipole chicanes or arc compressors.

The different types of compression scheme have similar projected energy spread because the energy spread is dominated by the chirp that is induced for the bunch compression process. The fact that the different compression schemes lead to comparable energy spread means that each scheme will lead to FEL pulses with similar relative bandwidth \cite{saldinCoherencePropertiesRadiation2008}. Energy spread, and therefore relative FEL bandwidth, can be reduced using corrugated de-chirpers to remove the negative chirp from chicane compression schemes \cite{baneCorrugatedPipeBeam2012}, or dielectric lined waveguides to remove the positive chirp from arc compression schemes \cite{paceySimulationStudiesDielectric2018}.

The longitudinal phase space distributions of electron bunches after the second bunch compressor from the arc, symmetric C-chicane and five-dipole chicane compression schemes are shown in Fig.~\ref{fig:sxl_ps}. The longitudinal phase space of the electron bunch from the second arc compression scheme (arc-2 in Table~\ref{tab:sxl_100pc_bunch_properties}) is not shown.

The slice properties (current, emittance and energy spread) were calculated for a slice with a duration that is on the order of the cooperation length ($L_c \approx 0.13\,\si{\femto\second}$), which characterises the length over which electrons can interact with each other in an FEL \cite{pellegriniPhysicsXRayFreeElectron2016,coffeeDevelopmentUltrafastCapabilities2019}. As the longitudinal phase space distributions of the $10\,\si{\pico\coulomb}$ electron bunches from the chicane compression schemes fold over, electrons in the longitudinal slices that have a large deviation from the median energy of the slice are excluded from the calculations of slice emittance and slice energy spread.

For the $100\,\si{\pico\coulomb}$ bunch charge case, electron bunches from the five-dipole chicane and arc compression schemes have roughly constant slice emittance along the bunch, whereas in the symmetric C-chicane scheme the slice emittance is increased for slices towards the head and tail of the bunch (Fig.~\ref{fig:sxl_slice}, bottom left). Only the arc compression scheme preserves slice emittance along the bunch for the $10\,\si{\pico\coulomb}$ bunch charge case (Fig.~\ref{fig:sxl_slice}, bottom right). This is because the CSR from the current spike at the tail of the electron bunches from chicane compression schemes interacts with particles in the middle and at the head of the bunch and degrades the slice emittance in these regions.

In the chicane compression schemes, the energy spread of the longitudinal slices towards the head and tail are much larger than in the longitudinal slices at the core of the bunch, while the opposite is true for electron bunches from arc compression schemes (Fig.~\ref{fig:sxl_slice}, middle row). The increased slice energy spread is due to the strong local compression factors and the mixing of electrons between neighbouring slices that occurs as a result.

The most significant differences in the longitudinal phase space distributions and bunch slice properties observed here, including the opposite sign for the chirp, higher-order components of the chirp, and the shape of the current profile, are dominated by the type of compression (chicane-like \emph{vs} arc-like), and the method used to linearise the longitudinal phase space distribution (magnetic/optical \emph{vs} harmonic RF).

\subsection{Free Electron Laser Performance}

Free electron laser performance in each case is evaluated using \texttt{GENESIS} simulations \cite{reicheGENESIS13Fully1999}.  The simulations use an undulator with period $\lambda_w = 4\,\si{\centi\metre}$ and on-axis magnetic field $B_0 = 0.33\,\si{\tesla}$ to generate FEL radiation with wavelength $\lambda_r = 1\,\si{\nano\meter}$ (critical photon energy $E_{ph} = 1.24\,\si{\kilo\electronvolt}$) from a $3\,\si{\giga\electronvolt}$ bunch.  The undulator is composed of $20$ modules with an active length of $2\,\si{\metre}$ per module ($40\,\si{\metre}$ total). The modules are separated by drift spaces with a quadrupole at the centre to provide periodic focusing.

\begin{figure}[t]
	\includegraphics[width=0.95\linewidth]{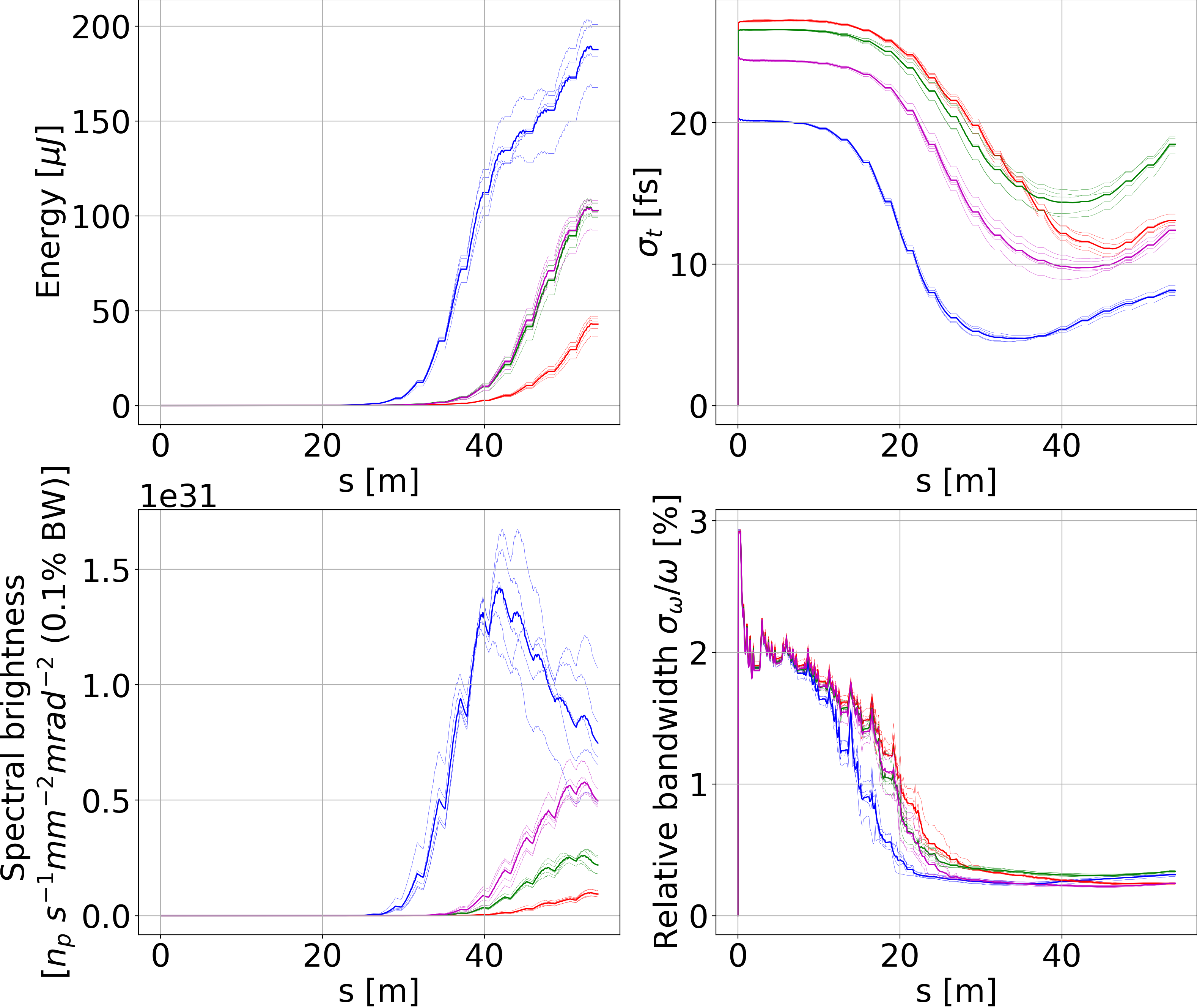}
	\caption{The FEL pulse energy (top left), duration (top right), spectral brightness (bottom left), and relative bandwidth (bottom right) as a function of distance $s$ along the undulator. Results from \texttt{GENESIS} simulations using $100\,\si{\pico\coulomb}$ bunches are shown, with bunches from the arc (blue), symmetric C-chicane (red), five-dipole chicane (green) and arc-2 (magenta) compression schemes. The thick solid lines shows the average of $5$ random shots.}
	\label{fig:sxl_100pc_genesis_undulator}
\end{figure}

\begin{figure}[t]
	\includegraphics[width=0.95\linewidth]{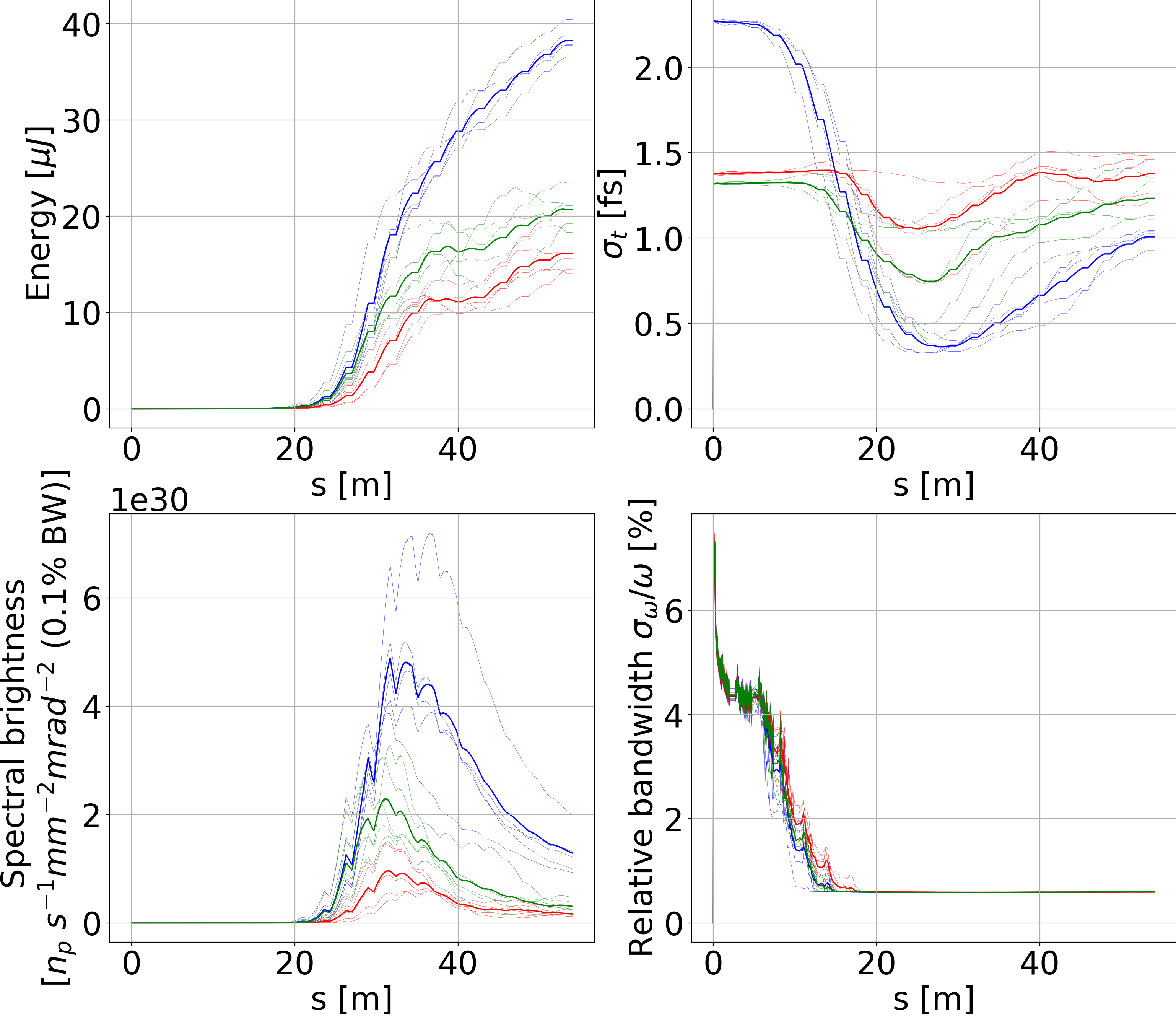}
	\caption{The FEL pulse energy (top left), duration (top right), spectral brightness (bottom left), and relative bandwidth (bottom right) as a function of distance $s$ along the undulator. Results from \texttt{GENESIS} simulations using a $10\,\si{\pico\coulomb}$ electron bunch are shown, with bunches from the arc (blue), symmetric C-chicane (red) and five-dipole chicane (green) compression schemes. The thick solid line shows the average of $5$ random shots.}
	\label{fig:sxl_10pc_genesis_undulator}
\end{figure}

Significant FEL properties (pulse energy, relative bandwidth, pulse duration and spectral brightness) as functions of distance along the undulator are shown for each compression scheme in Figs.~\ref{fig:sxl_100pc_genesis_undulator} (10\,\si{\pico\coulomb}) and \ref{fig:sxl_10pc_genesis_undulator} (100\,\si{\pico\coulomb}). The large peak currents and small slice emittances in electron bunches from the arc compression scheme result in significantly higher FEL pulse energies than is the case for the other compression schemes. The FEL pulse energy from the symmetric C-chicane is limited by the large projected and slice emittances of electron bunches in that case.

\begin{figure}[t]
	\includegraphics[width=0.95\linewidth]{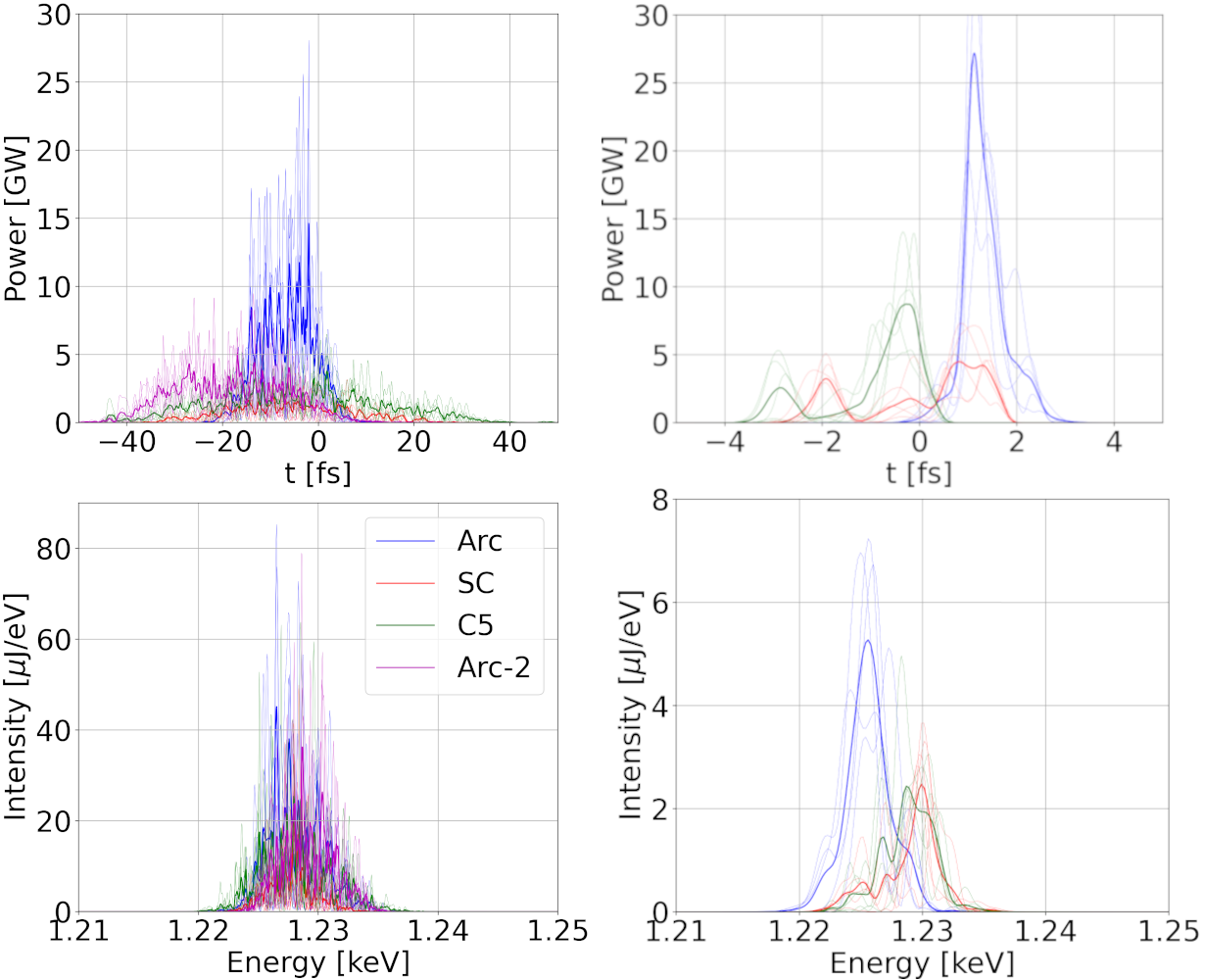}
	\caption{The longitudinal power profiles (top) and energy spectra (bottom) of FEL pulses generated by $100\,\si{\pico\coulomb}$ (left) and $10\,\si{\pico\coulomb}$ (right) electron bunches from the arc (blue), symmetric C-chicane (red), five-dipole chicane (green), and arc-2 (magenta) compression schemes. The thick solid line shows the average of a number of random shots.}
	\label{fig:sxl_genesis_pulse}
\end{figure}

The longitudinal power profiles and energy spectra of FEL pulses at peak spectral brightness are shown in Fig.~\ref{fig:sxl_genesis_pulse}.  For $10\,\si{\pico\coulomb}$ bunch charge, FEL pulses generated by electron bunches from the arc compression scheme have a single spike in the longitudinal power profile due to the ultra-short bunch length. Single spikes are also observed in FEL pulses generated by ultra-short electron bunches in Ref.~\cite{reicheDevelopmentUltrashortPulse2008}. The chicane compression schemes produce FEL radiation with longer pulse durations, and with longitudinal power profiles that have two spikes, coinciding with spikes in the current profiles of the electron bunches. The combination of high pulse energy and short pulse duration in the arc compression scheme results in a higher peak spectral brightness than is the case for FEL pulses generated by electron bunches from the other compression schemes.The relative bandwidths of the FEL pulses are broadly similar for each bunch charge in each compression scheme: this is because electron bunches in each case have similar projected energy spreads (see Tables~\ref{tab:sxl_100pc_bunch_properties} and \ref{tab:sxl_10pc_bunch_properties}).
%see Tables~\ref{tab:sxl_100pc_fel_properties} and \ref{tab:sxl_10pc_fel_properties}). 

Significant properties of the FEL pulses at peak spectral brightness for each compession scheme are summarised in
Tables~\ref{tab:sxl_100pc_fel_properties} ($100\,\si{\pico\coulomb}$) and \ref{tab:sxl_10pc_fel_properties} ($10\,\si{\pico\coulomb}$).

%The single central spike in the current profile of the bunch from the arc compression scheme (see Figure~\ref{fig:sxl_slice}) results in longitudinal power profiles of the FEL pulses that also have a central spike. The FEL pulses from the arc compression scheme have shorter pulse durations than those from either of the chicane compression schemes. 

\begin{table}[t]
	\caption{FEL pulse properties at peak spectral brightness for each compression scheme using a bunch charge of $100\,\si{\pico\coulomb}$.}
	\begin{ruledtabular}
		\begin{tabular}{l c c c c}
			Pulse properties & Arc & SC & C5 & Arc-2 \\
			\hline
            Spectral brightness & \multirow{2}{2em}{14.2} & \multirow{2}{2em}{0.99} & \multirow{2}{2em}{2.60} & \multirow{2}{2em}{5.78} \\
            {[$10^{30}n_p / \si{\second} / \si{\milli\metre}^2 / \si{\milli\radian}^2 / 0.1 \si{\percent} BW$]} \\
            %\textrm{Spectral brightness [$10^{30}\times$} & & & & \\
			%\textrm{$\si{n_p / \second / \milli\metre^2 / \milli\radian^2 / 0.1 \percent BW}$]} & 14.19 & 0.99 & 2.60 & 5.78 \\
			Energy [\si{\micro\joule}] & 132.8 & 43.1 & 102.9 & 102.8 \\
			Pulse duration $\sigma_t$ [\si{\femto\second}] & 5.81 & 13.06 & 17.68 & 11.90 \\
			Relative bandwidth $\sigma_{\omega}/\omega$ [\si{\percent}] & 0.27 & 0.24 & 0.33 & 0.24 \\
			Peak power [\si{\giga\watt}] & 14.64 & 1.95 & 3.81 & 5.14 \\
			Peak intensity [\si{\micro\joule / \electronvolt}] & 45.18 & 20.86 & 25.39 & 36.28 \\
		\end{tabular}
	\end{ruledtabular}
	\label{tab:sxl_100pc_fel_properties}
\end{table}

\begin{table}[t]
	\caption{FEL pulse properties at peak spectral brightness for each compression scheme using a bunch charge of $10\,\si{\pico\coulomb}$.}
	\begin{ruledtabular}
		\begin{tabular}{l c c c}
			Pulse properties & Arc & SC & C5 \\
			\hline
            Spectral brightness & \multirow{2}{2em}{4.89} & \multirow{2}{2em}{0.96} & \multirow{2}{2em}{2.29} \\
            {[$10^{30}n_p / \si{\second} / \si{\milli\metre}^2 / \si{\milli\radian}^2 / 0.1 \si{\percent} BW$]} \\
			Pulse energy [\si{\micro\joule}] & 18.07 & 6.92 & 11.07 \\
			Pulse duration $\sigma_t$ [\si{\femto\second}] & 0.42 & 1.18 & 0.90 \\
			Relative bandwidth $\sigma_{\omega}/\omega$ [\si{\percent}] & 0.58 & 0.59 & 0.59 \\
			Peak power [\si{\giga\watt}] & 27.19 & 4.50 & 8.74 \\
			Peak intensity [\si{\micro\joule / \electronvolt}] & 5.27 & 2.48 & 2.44 \\
		\end{tabular}
	\end{ruledtabular}
	\label{tab:sxl_10pc_fel_properties}
\end{table}

\subsection{Compression Factor Scan}

\begin{figure}[ht]
	\includegraphics[width=0.495\linewidth]{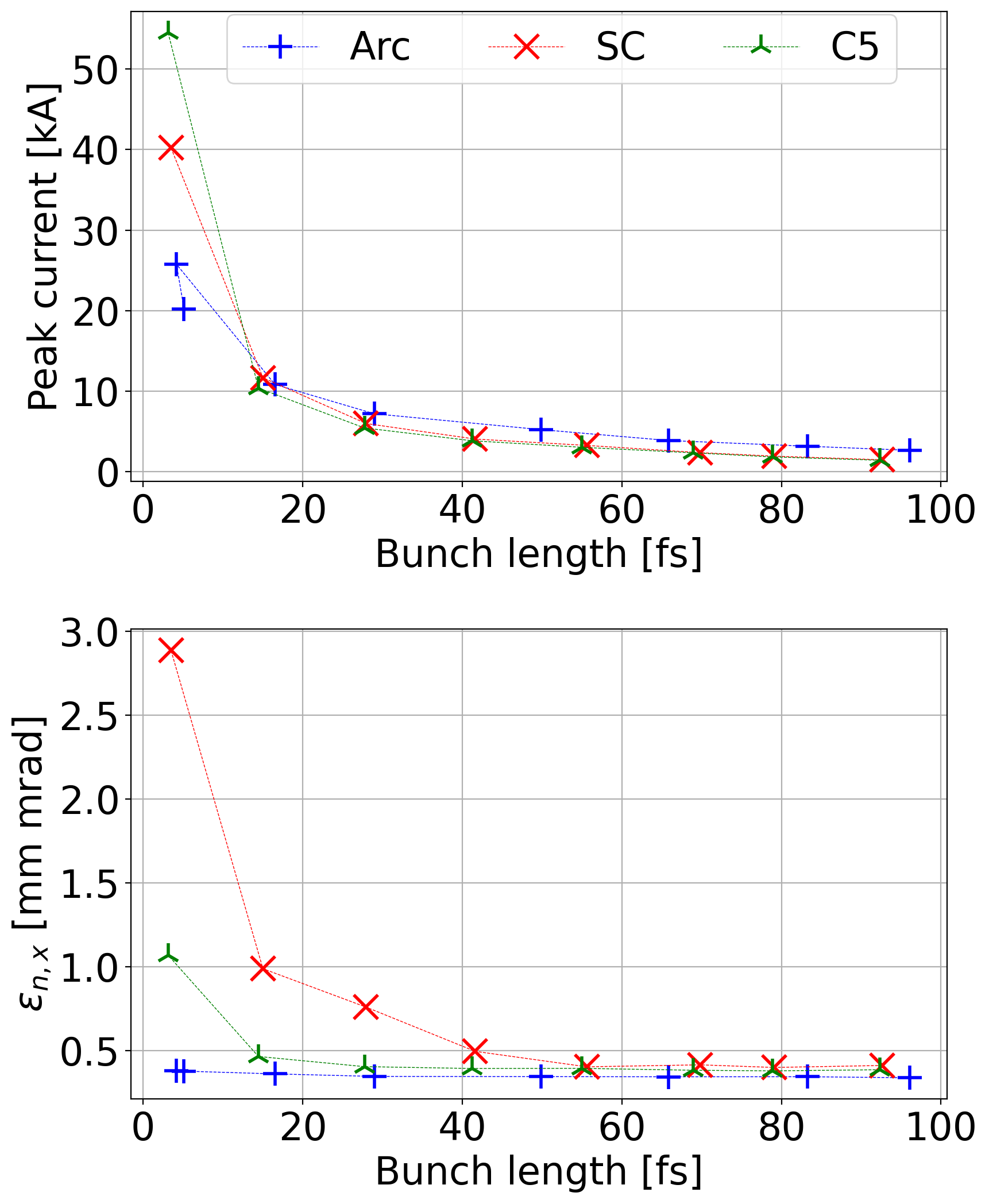}
	\includegraphics[width=0.485\linewidth]{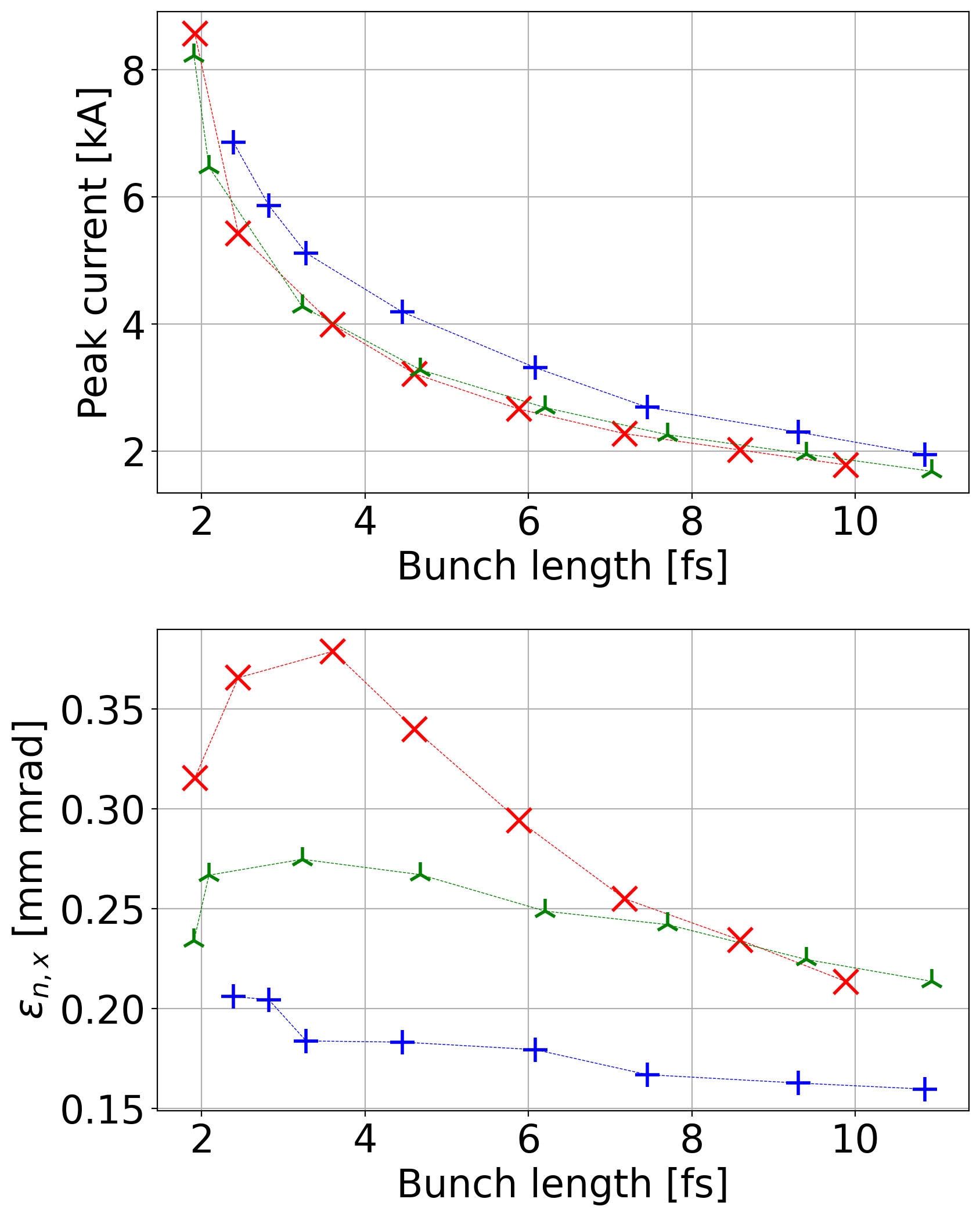}
	\caption{Peak current (top) and horizontal emittance of peak current slice (bottom) as a function of bunch length from compression scans of each compression scheme using bunch charges of $100\,\si{\pico\coulomb}$ (left) and $10\,\si{\pico\coulomb}$ (right). The bunch length is scanned by varying RF phase in linac 2 to control the compression factor in the second bunch compressor.} 
\label{fig:sxl_pc_senx}
\end{figure}

Compression factor, i.e. the ratio of initial bunch length to final bunch length, is scanned for each scheme to investigate the emittance growth due to CSR at stronger compression factors and larger peak currents. For each of the three designs shown in Fig.~\ref{fig:all_layouts}, the compression factor in the second bunch compressor was varied by changing the RF phase in the linac between the first and second bunch compressor (linac 2) while simultaneously adjusting the RF voltage to maintain constant energy at the bunch compressor. The final bunch lengths approximately cover a range $5\,\si{\femto\second} < \textrm{FWTM} < 100\,\si{\femto\second}$ for a bunch charge of $100\,\si{\pico\coulomb}$, and a range $2\,\si{\femto\second} < \textrm{FWTM} < 10\,\si{\femto\second}$ for a bunch charge of $10\,\si{\pico\coulomb}$. Fig.~\ref{fig:sxl_pc_senx} shows the peak current and the emittance of the peak current slice as functions of bunch length for each case.

At longer bunch lengths, the peak currents of electron bunches from the arc compression schemes are typically larger than those from the chicane compression schemes, as a result of the single-spike current profiles. However, at shorter bunch lengths and stronger compression factors, specifically for a bunch charge of $100\,\si{\pico\coulomb}$, the electron bunches from chicane compression schemes develop strong non-linearities in the longitudinal phase space distribution in which electrons at the head and tail fold over the core of the bunch \cite{charlesCausticBasedApproachUnderstanding2016,charlesCurrentHornSuppressionReduced2017}. The folds in the longitudinal phase space form sharp current spikes resulting in larger peak currents than those from the arc compression scheme. 

As seen in the bottom plots in Fig.~\ref{fig:sxl_pc_senx}, the sharp current spikes that form at short bunch lengths due to non-linearities in the longitudinal phase space distribution from the symmetric C-chicane, and to a lesser extent the five-dipole chicane, typically result in significant emittance growth. The severity of the emittance growth can be seen when the slice emittances in the symmetric C-chicane are compared to those from the arc compression scheme: in the latter case, the slice emittance is well preserved because the longitudinal phase space distribution does not fold over.

\begin{figure}[t]
	\includegraphics[width=0.49\linewidth]{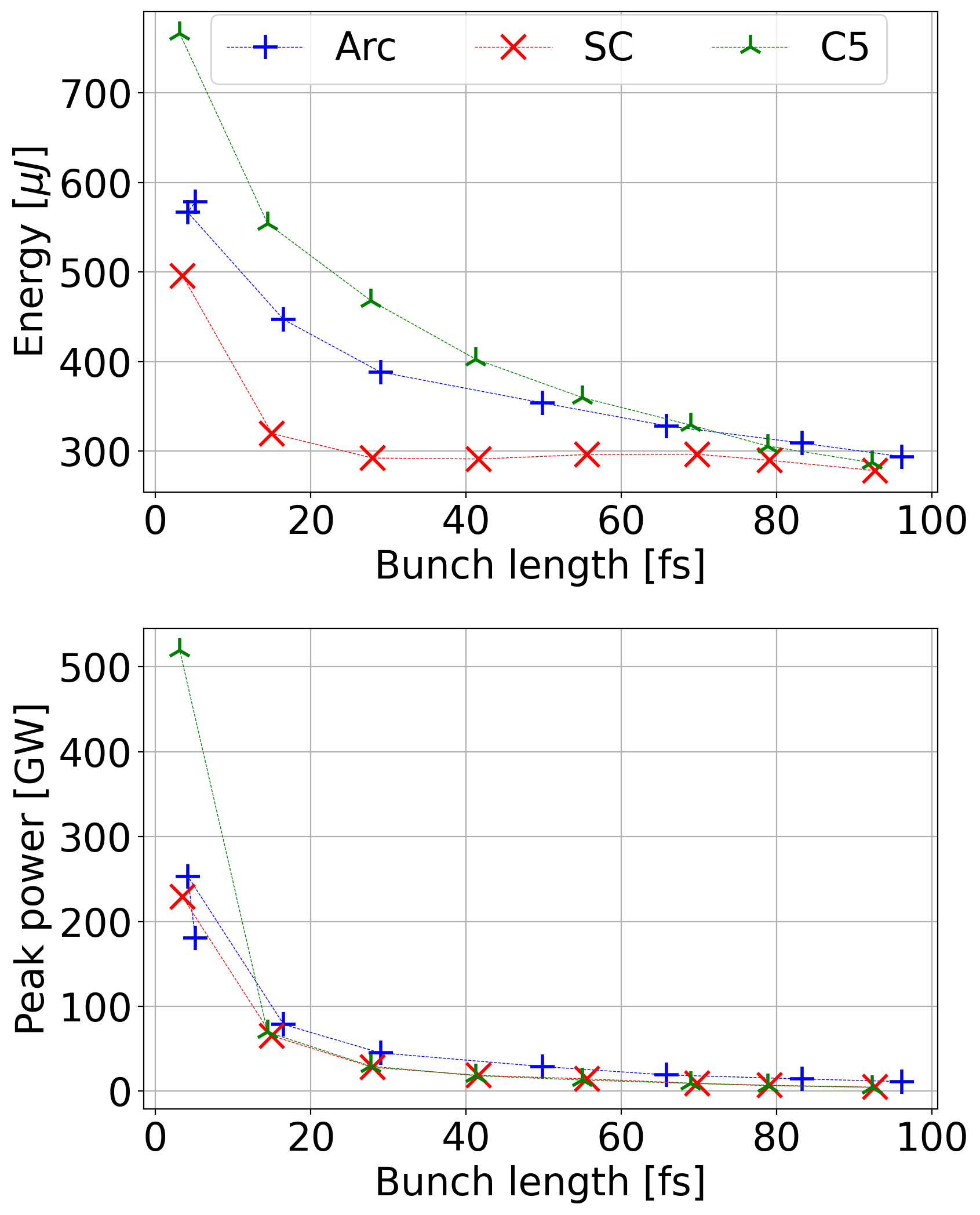}
	\includegraphics[width=0.49\linewidth]{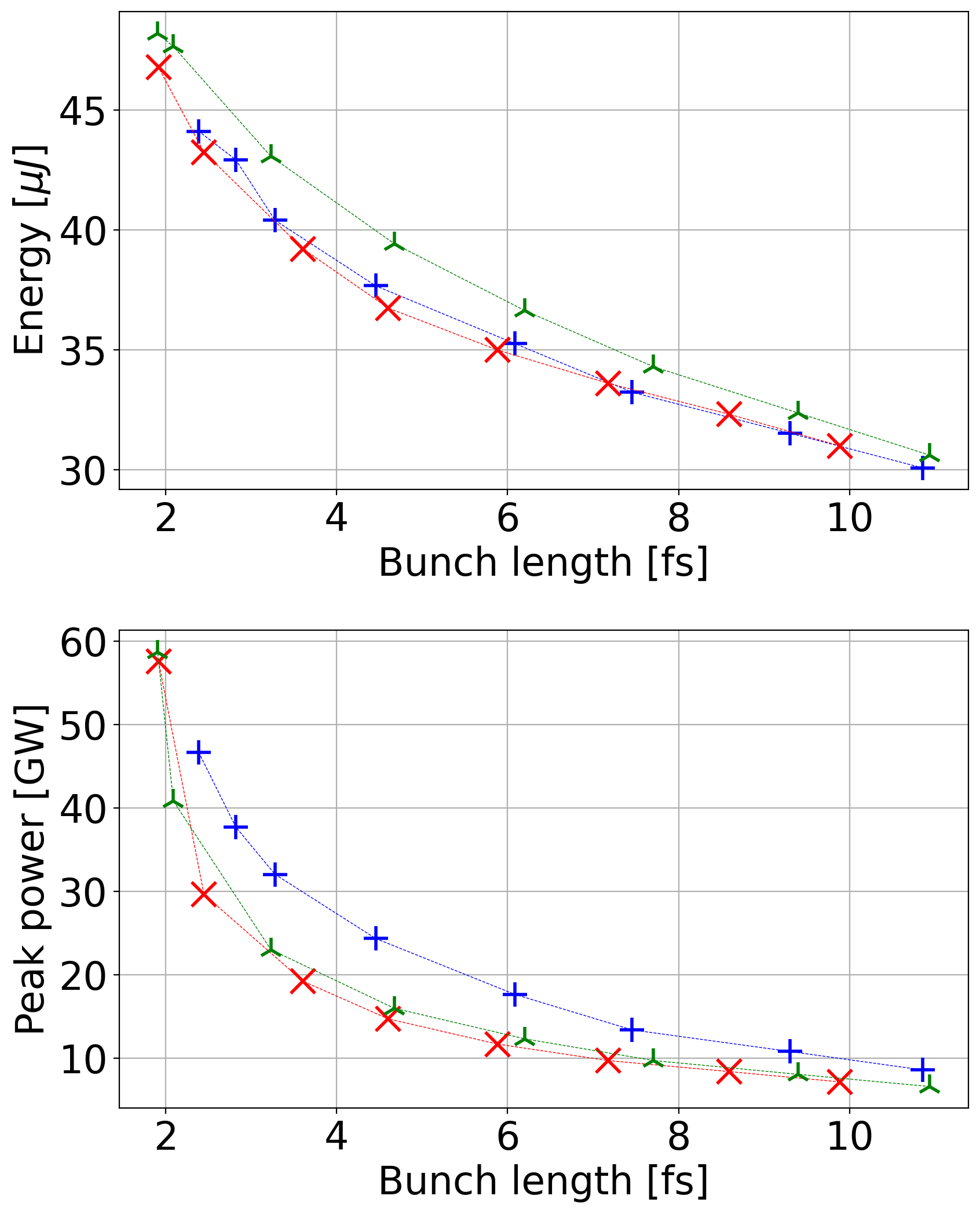}
	\caption{Pulse energy (top) and peak power (bottom) as a function of bunch length from compression scans of each compression scheme using bunch charges of $100\,\si{\pico\coulomb}$ (left) and $10\,\si{\pico\coulomb}$ (right). Pulse energy and peak power are calculated analytically from the slice electron bunch properties. The bunch length is scanned by varying RF phase in linac 2 to control the compression factor in the second bunch compressor.}
	\label{fig:sxl_mingxie}
\end{figure}

Using theory from Ref.~\cite{mingxieDesignOptimizationXray1995}, the slice properties of the electron bunches are used to make analytical estimates of FEL peak power and pulse energy. The results are shown in Fig.~\ref{fig:sxl_mingxie}.
The estimated pulse energies for FEL pulses generated by electron bunches from the five-dipole chicane are generally larger than is the case in the other compression schemes, over the range of bunch lengths studied. This higher pulse energy can be explained by the relatively small slice emittance, the small slice energy spread, and the moderately high current along the whole bunch, which allows all slices to contribute to lasing. Electron bunches from the arc compression scheme also have small slice emittance and energy spread along the whole bunch, but the low current at the head and tail means that these parts of the bunch do not lase as efficiently. Nevertheless, the high current at the core of the bunches from arc compression schemes generates large peak power.

\subsection{Charge Sensitivity}

Charge jitter occurs in an accelerator due to random variations in the electron source. Variations in bunch charge will result in changes to the electron bunch properties, which will in turn lead to variations in the output of an FEL. %It is therefore important to understand the effect that variations in bunch charge might have on the properties of an electron bunch.
To characterise the sensitivity of significant bunch properties (bunch length, peak current, emittance and energy spread) to variations in bunch charge, simulations have been performed in which the bunch charge is varied over a range of about $\pm 10\,\si{\percent}$.  Bunches are tracked from the start of the first linac, and properties are calculated after the second bunch compressor. Note that as the charge variations are introduced at the start of the first linac, the effect of charge variations at low energies ($<100\,\si{\mega\electronvolt}$) in the injector are not included.  The sensitivity of a given parameter is calculated from the gradient of a linear fit to the change in that parameter with respect to the change in the bunch charge around the nominal bunch charge.  For a nominal bunch charge of $100\,\si{\pico\coulomb}$, the bunch charge is scanned from $90\,\si{\pico\coulomb}$ to $110\,\si{\pico\coulomb}$; results are shown in Table~\ref{tab:sxl_100pc_charge_jitter}.  For $10\,\si{\pico\coulomb}$ nominal bunch charge, the charge is scanned from $9\,\si{\pico\coulomb}$ to $11\,\si{\pico\coulomb}$; results are shown in Table~\ref{tab:sxl_10pc_charge_jitter}.

%For a nominal bunch charge of $100\,\si{\pico\coulomb}$, the bunch charge is scanned from $90$ to $110\,\si{\pico\coulomb}$. The sensitivities of the electron bunch properties to bunch charge are from the gradient of a linear fit to the data, they are summarised in Table~\ref{tab:sxl_100pc_charge_jitter}.

%For the compression scheme using a nominal bunch charge of $10\,\si{\pico\coulomb}$, the initial bunch charge at the start of linac 1 is scanned from $9$ to $11\,\si{\pico\coulomb}$. The sensitivities of the electron bunch properties to bunch charge for each compression scheme are summarised in Table~\ref{tab:sxl_10pc_charge_jitter}. 

\begin{table}[t]
	\renewcommand{\arraystretch}{1.4}
	\caption{Relative sensitivity of electron bunch properties to variations in bunch charge for each bunch compression scheme with a nominal bunch charge of $100\,\si{\pico\coulomb}$. The sensitivity values are the gradients of a linear fit.}
	\begin{ruledtabular}
		\begin{tabular}{l c c c}
			Charge sensitivity & Arc & SC & C5 \\
			\hline
			%$\frac{\Delta FWTM_t}{\Delta Q}$ [$\si{\femto\second/\pico\coulomb}$] & -3.17 & 1.35 & 1.32 \\
			${\Delta \textrm{FWTM} / \textrm{FWTM}_{\mathrm{nom}}}/{\Delta Q}$ [$\si{\percent/\pico\coulomb}$] & -3.17 & 1.33 & 1.32 \\
			%$\frac{\Delta I_{peak}}{\Delta dQ}$ [$\si{\kilo\ampere/\pico\coulomb}$] & 0.15 & -0.02 & -0.02 \\
			${\Delta I_{\mathrm{peak}} / I_{\mathrm{peak},\mathrm{nom}}}/{\Delta Q}$ [$\si{\percent/\pico\coulomb}$] & 6.22 & -1.38 & -1.54 \\
			%$\frac{\Delta \varepsilon_{n,x}}{\Delta dQ}$ [$\si{\milli\metre \ \milli\radian/\pico\coulomb}$] & $1.31\times10^{-3}$ & $1.28\times10^{-3}$ & $0.08\times10^{-3}$ \\
			${\Delta \varepsilon_{n,x} / \varepsilon_{n,x,\mathrm{nom}}}/{\Delta Q}$ [$\si{\percent/\pico\coulomb}$] & 0.31 & 0.15 & 0.02 \\
			%$\frac{\Delta \sigma_{\delta}}{\Delta dQ}$ [$\si{\percent/\pico\coulomb}$] & $0.38\times10^{-3}$ & $-0.27\times10^{-3}$ & $-0.29\times10^{-3}$ \\
			${\Delta \sigma_{\delta} / \sigma_{\delta,\mathrm{nom}}}/{\Delta Q}$ [$\si{\percent/\pico\coulomb}$] & 0.18 & -0.15 & -0.15 \\
		\end{tabular}
	\end{ruledtabular}
	\label{tab:sxl_100pc_charge_jitter}
\end{table}

\begin{table}[t]
	\renewcommand{\arraystretch}{1.4}
	\caption{Relative sensitivity of electron bunch properties to variations in bunch charge for each bunch compression scheme with a nominal bunch charge of $10\,\si{\pico\coulomb}$. The sensitivity values are the gradients of a linear fit.}
	\begin{ruledtabular}
		\begin{tabular}{l c c c}
			Charge sensitivity & Arc & SC & C5 \\
			\hline
			%$\frac{\Delta FWTM_t}{\Delta Q}$ [$\si{\femto\second/\pico\coulomb}$] & -3.17 & 1.35 & 1.32 \\
			${\Delta \textrm{FWTM} / \textrm{FWTM}_{\mathrm{nom}}}/{\Delta Q}$ [$\si{\percent/\pico\coulomb}$] & -45.77 & 22.13 & 23.99 \\
			%$\frac{\Delta I_{peak}}{\Delta dQ}$ [$\si{\kilo\ampere/\pico\coulomb}$] & 0.15 & -0.02 & -0.02 \\
			${\Delta I_{\mathrm{peak}} / I_{\mathrm{peak},\mathrm{nom}}}/{\Delta Q}$ [$\si{\percent/\pico\coulomb}$] & 46.65 & -6.03 & -6.12 \\
			%$\frac{\Delta \varepsilon_{n,x}}{\Delta dQ}$ [$\si{\milli\metre \ \milli\radian/\pico\coulomb}$] & $1.31\times10^{-3}$ & $1.28\times10^{-3}$ & $0.08\times10^{-3}$ \\
			${\Delta \varepsilon_{n,x} / \varepsilon_{n,x,\mathrm{nom}}}/{\Delta Q}$ [$\si{\percent/\pico\coulomb}$] & 6.90 & 4.55 & 6.06 \\
			%$\frac{\Delta \sigma_{\delta}}{\Delta dQ}$ [$\si{\percent/\pico\coulomb}$] & $0.38\times10^{-3}$ & $-0.27\times10^{-3}$ & $-0.29\times10^{-3}$ \\
			${\Delta \sigma_{\delta} / \sigma_{\delta,\mathrm{nom}}}/{\Delta Q}$ [$\si{\percent/\pico\coulomb}$] & 0.62 & 1.05 & 1.33 \\
		\end{tabular}
	\end{ruledtabular}
	\label{tab:sxl_10pc_charge_jitter}
\end{table}

The peak current and bunch length of the electron bunch are more sensitive to variations in charge for the arc compression scheme compared with the chicane compression schemes due to a self-stabilising effect in the chicane compression schemes. The stabilising effect in a chicane compression scheme can be explained as follows: a smaller bunch charge leads to weaker short range wakefields and a larger chirp, which in turn leads to a stronger compression factor in the first chicane bunch compressor. The stronger compression factor in the first bunch compressor leads to a shorter bunch that produces stronger short range wakefields, resulting in a smaller chirp and a weaker compression factor in the following chicane bunch compressor. Additionally, the arc compression scheme has a higher peak current at the nominal charge for the $100\,\si{\pico\coulomb}$ case, which generates comparatively stronger short range wakefields and increases charge sensitivity.

Despite the stronger sensitivity of peak current to variations in charge for the $10\,\si{\pico\coulomb}$ arc compression case, the sensitivity of projected emittance to variations in bunch charge is similar in all three bunch compression schemes. This could be due to the non-linear longitudinal phase space distributions from the chicane compression schemes which produce more significant CSR effects than the single-spiked current profile in the arc scheme. For the $100\,\si{\pico\coulomb}$ case, sensitivity of emittance to bunch charge is smaller for the five-dipole chicane compression scheme than for the others. This is a combined result of the weak sensitivity of peak current to bunch charge variations and effective mitigation of emittance growth due to CSR.

\section{Application to 8 GeV Hard X-ray FEL: UK-XFEL}
\label{sec:bc_comp_ukxfel}

In this section, the three bunch compressors discussed in Sec.~\ref{sec:bc_comp_types} are compared in the context of possible application to a hard X-ray ($4 - 100 \ \si{\kilo\electronvolt}$) FEL, driven by a super-conducting linac. This case is motivated by the UK X-ray Free Electron Laser project \cite{UKXFELScienceCase2020}, which proposes a high repetition-rate linac to drive multiple beamlines operating simultaneously in different modes, such as SASE, self-seeded and HB-SASE. Providing this capability will require the ability to achieve bunch-by-bunch variations in the electron bunch properties, particularly charge, current, bunch length and energy spread. Optimal performance of each FEL beamline will depend on precise control of all these properties. %Emittance preservation is vital for operating a high performance FEL in the hard X-ray regime, so UK-XFEL requires a bunch compression scheme that can preserve the transverse emittance, while providing flexibility in the longitudinal bunch properties.

UK-XFEL proposes a $1.3\,\si{\giga\hertz}$ super-conducting linac to accelerate electron bunches to $8\,\si{\giga\electronvolt}$. The necessary peak currents will be achieved using a two-stage compression scheme. In order to make a comparison between the three bunch compressor configurations considered in the current work, the first-order longitudinal dispersion and beam energy of each bunch compressor are based on the SHINE parameters \cite{zhaoSCLF8GeVCW2018,fengReviewFullyCoherent2018}. Bunch compression occurs at $270\,\si{\mega\electronvolt}$ and $2.1\,\si{\giga\electronvolt}$. 

The geometry of the symmetric C-chicane bunch compressor (bend angle, dipole lengths and drift lengths) is optimised for minimising CSR-induced emittance growth. The five-dipole chicane uses dipoles of the same length, and the geometry is similarly optimised to minimise CSR-induced emittance growth.

\begin{figure}[t]
	%\centering
	\includegraphics[width=0.95\linewidth]{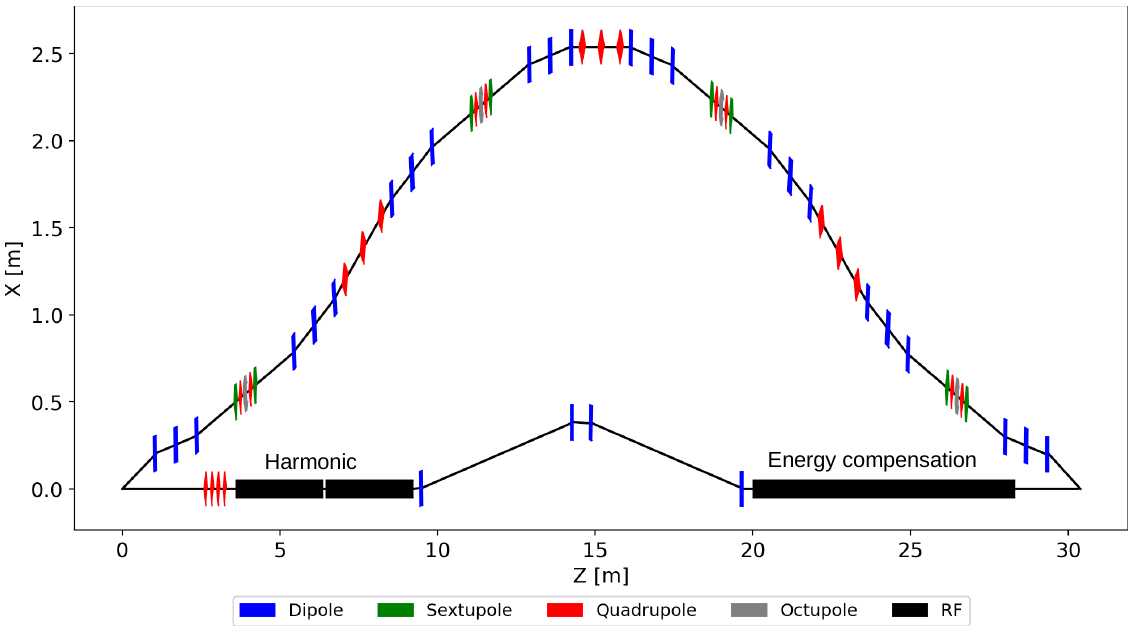}
	\caption{Layout of the first arc bunch compressor bypassing the symmetric C-chicane, harmonic cavity and energy compensating cavity. Note: additional space would be required for fast kickers and drifts to separate bunches intended for the arc bunch compressor.}
	\label{fig:ukxfel_bc_layout}
\end{figure}

The achromats of the arc bunch compressors are based on the arc-like variable bunch compressor design published in~\cite{williamsArclikeVariableBunch2020,dixonReductionArrivalTime2024}. The arc compressors considered for MAX IV consisted of two achromats; for UK-XFEL, the arc bunch compressors consist of four six-dipole achromats, where the final two achromats are a mirrored version of the first two. The additional achromats bring the exit of the UK-XFEL bunch compressor back on axis, so there is no horizontal translation of the downstream components (as was the case for MAX IV), and allows better control over chromatic effects. Some modifications are made to the arc bunch compressor design from~\cite{williamsArclikeVariableBunch2020,dixonReductionArrivalTime2024}: dipole lengths are the same as in the chicane bunch compressors, and there is a small reduction in drift lengths within the dipole triplets. A schematic of the first arc bunch compressor, which bypasses the first symmetric C-chicane bunch compressor, is shown in Fig.~\ref{fig:ukxfel_bc_layout}.

The four-achromat bunch compressor design was chosen as it is able to bypass a symmetric C-chicane bunch compressor (or five-dipole chicane bunch compressor), including the harmonic RF cavity and an energy compensating RF cavity. This design could allow either bunch compressor to be used to compress electron bunches in UK-XFEL in order to generate bunch-by-bunch variations in the longitudinal electron beam properties. %Bunch-by-bunch variations in the longitudinal electron bunch properties enables a wide range of FEL operating modes to be used simultaneously.
Other methods are being considered to achieve control of the longitudinal electron bunch properties on a bunch-by-bunch basis, such as: the use of fast kickers within a bunch compressor to change the longitudinal dispersion, and induce bunch-by-bunch variations in bunch length and peak current \cite{zhuFlexibleMultibunchlengthOperation2024}; the use of a normal-conducting cavity to alter the chirp of selected bunches \cite{zhangFastFlexibleControl2023}; the use of multiple double-bend achromats to delay some bunches relative to others, so that they arrive in the following RF cavities different voltages, which achieves a large degree of flexibility in the electron beam energy through the use of partial energy recovery \cite{yanMultiBeamEnergyOperationContinuousWave2019,wuMultiBeamEnergyControlUnit2025}.

\begin{figure}[t]
	%\centering
	\includegraphics[width=0.95\linewidth]{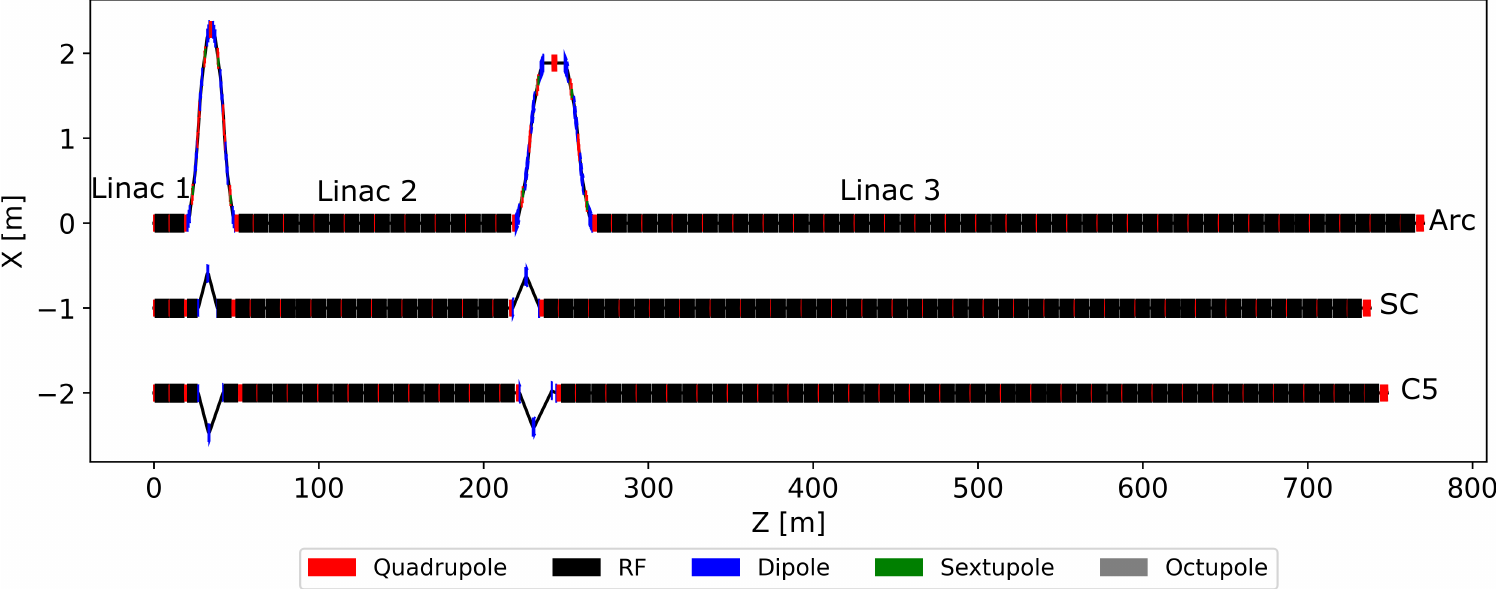}
	\caption{Layout of the linac with arc compressors (top), symmetric C-chicane compressors (middle) and five-dipole chicane compressors (bottom).}
	\label{fig:ukxfel_comp_layout}
\end{figure}

Figure~\ref{fig:ukxfel_comp_layout} shows the layout of the proposed UK-XFEL linac with the different bunch compressors used in this comparison. The total length of the linac (including bunch compressors) is slightly larger for the arc bunch compression scheme than for the chicane schemes; however, the difference in length between the longest and shortest design is no more than $32\,\si{\metre}$. The length of the first arc bunch compressor is comparable to that of the symmetric C-chicane (or five-dipole chicane), harmonic cavity and energy compensating cavity. The difference in length is mainly due to the second bunch compressor. 

As with the arc bunch compressors in Sec.~\ref{sec:bc_comp_maxiv}, the four-achromat arc bunch compressors used in this section produce stronger chromatic effects than either of the chicane bunch compressors. However, the impact of chromatic effects on the projected emittance is negligible, see Tables~\ref{tab:ukxfel_75pc_bunch_properties} and \ref{tab:ukxfel_300pc_bunch_properties}. Details of these studies are given in Ref.~\cite{dixonBeamDynamicsBunch2025}.

The bunch compression schemes are compared for two bunch charges and bunch lengths:
\begin{enumerate}
	\item[A.] $75\,\si{\pico\coulomb}$ bunch compressed to a final bunch length of $40\,\si{\femto\second}$.
	\item[B.] $300\,\si{\pico\coulomb}$ bunch compressed to a final bunch length of $100\,\si{\femto\second}$.
\end{enumerate}

Details of the specific RF phases and compression ratios used in each compression scheme can be found in Ref.~\cite{dixonBeamDynamicsBunch2025}. 

The initial particle distributions are generated in \texttt{ELEGANT} \cite{borlandElegantFlexibleSDDSCompliant2000}, assuming ideal Gaussian distributions in the three orthogonal 2D planes ($x-x'$, $y-y'$ and $t-\delta$) and no correlations between the coordinates in the three planes. The transverse emittance and energy spread used to generate the electron distributions are the statistical values calculated from electron bunches from simulations of an injector with a Very High Frequency (VHF) electron gun \cite{davutBalanceBunchCompression2025}. 

Tables~\ref{tab:ukxfel_75pc_bunch_properties} and \ref{tab:ukxfel_300pc_bunch_properties} show the electron bunch properties at the exit of linac 3 (the final accelerating section) for the $75\,\si{\pico\coulomb}$ and $300\,\si{\pico\coulomb}$ bunch charge cases. As in Sec.~\ref{sec:bc_comp_maxiv}, an additional arc compression case (arc-2), which produces bunches with a similar peak current to the electron bunches from the chicane compression schemes, has been included in the comparison.

\begin{table}[t]
	\caption{Electron bunch properties at the entrance of the FEL for each compression scheme using a bunch charge of $75\,\si{\pico\coulomb}$.}
	\begin{ruledtabular}
		\begin{tabular}{l c c c c}
			Bunch properties & Arc & SC & C5 & Arc-2 \\
			\hline
			Energy [\si{\giga\electronvolt}] & 8.00 & 8.00 & 8.00 & 8.00 \\
			Peak current [\si{\kilo\ampere}] & 5.24 & 2.78 & 2.24 & 2.27 \\
			Bunch length [\si{\femto\second}] & 46.00 & 40.04 & 41.48 & 82.31 \\
			$\varepsilon_{n,x/y}$ [\si{\milli\meter \milli\radian}] & 0.15/0.08 & 0.18/0.08 & 0.09/0.08 & 0.09/0.08 \\
			Energy spread [\si{\percent}] & 0.11 & 0.06 & 0.06 & 0.09 \\
		\end{tabular}
	\end{ruledtabular}
	\label{tab:ukxfel_75pc_bunch_properties}
\end{table}

\begin{table}[t]
	\caption{Electron bunch properties at the entrance of the FEL for each compression scheme using a bunch charge of $300\,\si{\pico\coulomb}$.}
	\begin{ruledtabular}
		\begin{tabular}{l c c c c}
			Bunch properties & Arc & SC & C5 & Arc-2 \\
			\hline
			Energy [\si{\giga\electronvolt}] & 8.00 & 8.00 & 8.00 & 8.00 \\
			Peak current [\si{\kilo\ampere}] & 7.69 & 5.20 & 5.64 & 5.21 \\
			Bunch length [\si{\femto\second}] & 100.63 & 100.19 & 99.60 & 148.24 \\
			$\varepsilon_{n,x/y}$ [\si{\milli\meter \milli\radian}] & 0.30/0.25 & 0.46/0.21 & 0.25/0.21 & 0.25/0.25 \\
			Energy spread [\si{\percent}] & 0.16 & 0.10 & 0.10 & 0.15 \\
		\end{tabular}
	\end{ruledtabular}
	\label{tab:ukxfel_300pc_bunch_properties}
\end{table}

\begin{figure}[t]
	\includegraphics[width=0.4975\linewidth]{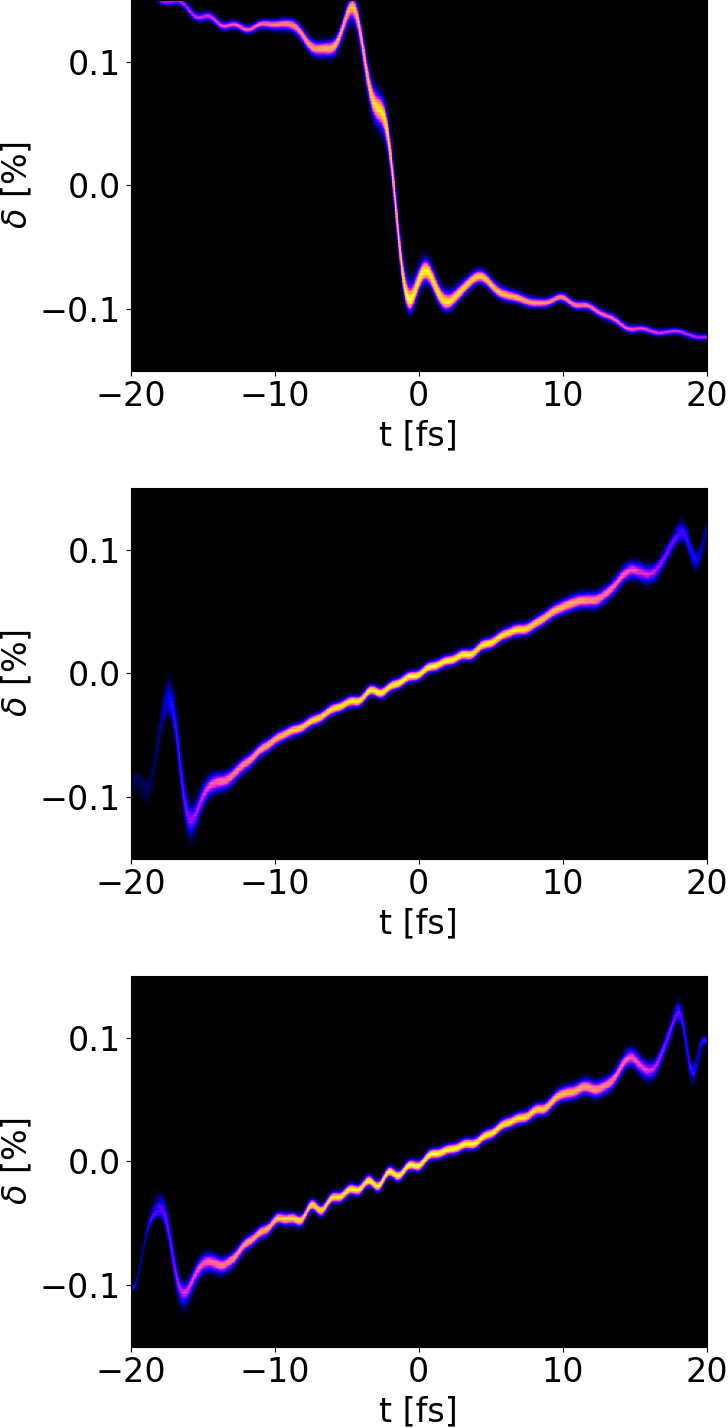}
	\includegraphics[width=0.4825\linewidth]{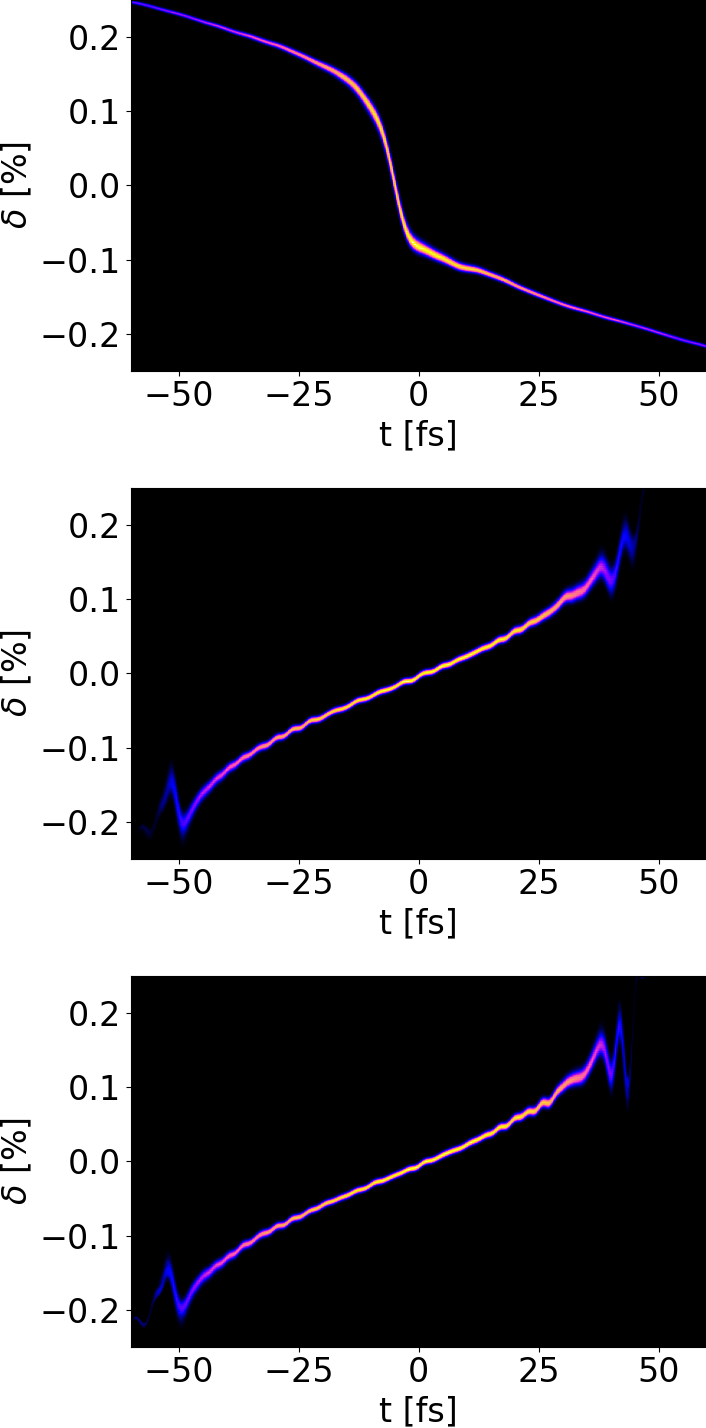}
	\caption{Longitudinal phase space distributions of electron bunches from arc (top), symmetric C-chicane (middle), and five-dipole chicane (bottom) compression schemes for bunch charges: $75\,\si{\pico\coulomb}$ (left) and  $300\,\si{\pico\coulomb}$ (right).}
	\label{fig:ukxfel_ps}
\end{figure}

\begin{figure}[t]
	\includegraphics[width=0.4975\linewidth]{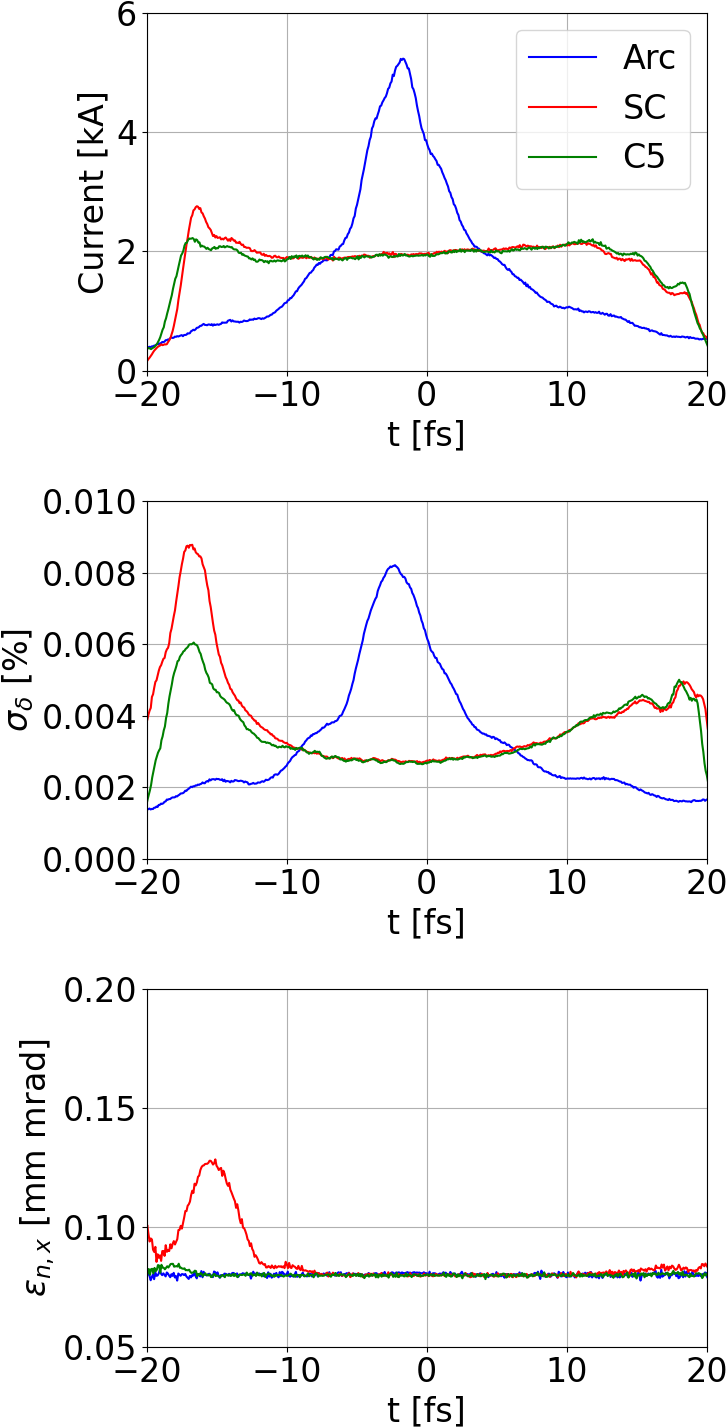}
	\includegraphics[width=0.4825\linewidth]{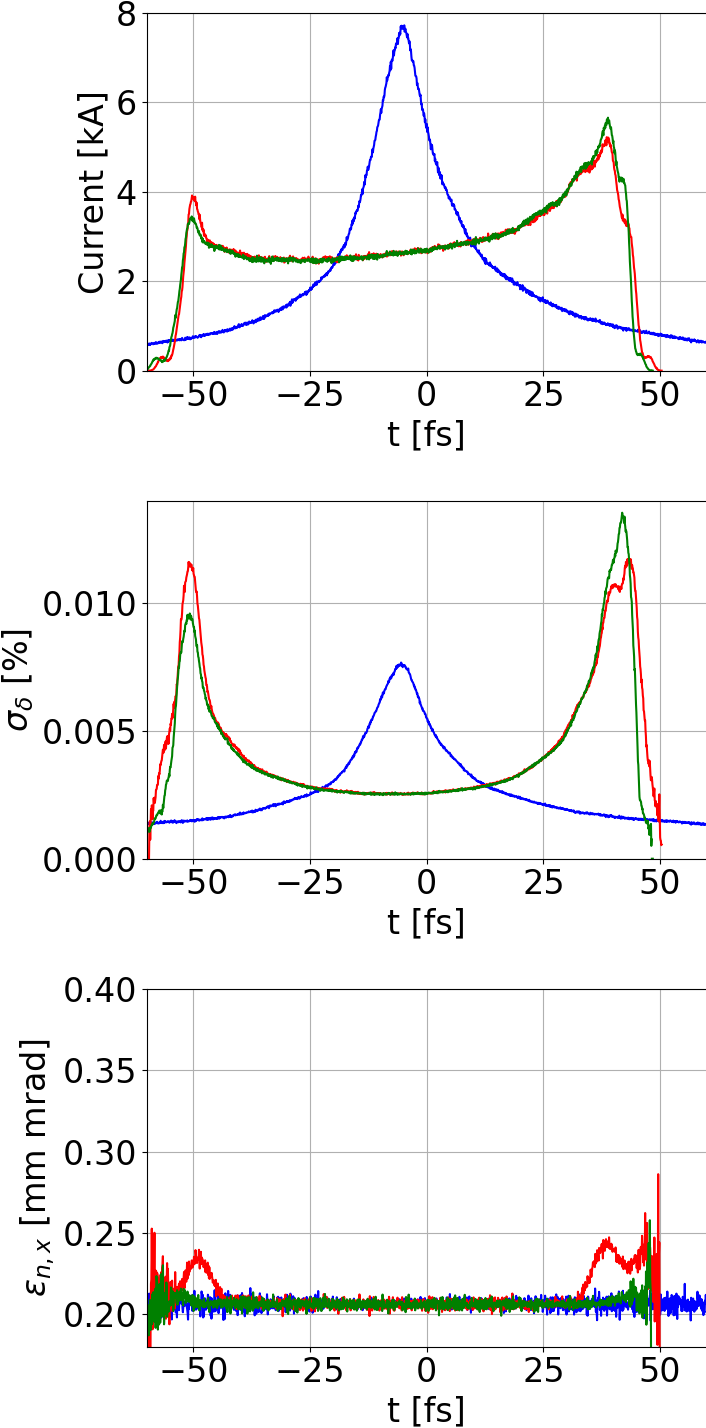}
	\caption{Current (top), slice energy spread (middle) and horizontal slice emittance (bottom) of electron bunches from arc (blue), symmetric C-chicane (red), and five-dipole chicane (green) compression schemes for bunch charges: $75\,\si{\pico\coulomb}$ (left) and  $300\,\si{\pico\coulomb}$ (right).}
	\label{fig:ukxfel_slice}
\end{figure}

Electron bunches compressed in the arc compression scheme have significantly higher peak currents (at equivalent bunch length) as a result of the strong local compression factor at the core of the bunch, which results from the chirp being enhanced by CSR and short range wakefields (see Fig.~\ref{fig:ukxfel_ps}). The strong chirp at the core of the electron bunch also produces a single peak in the current profile, unlike the uniform current profile of the electron bunches from the chicane compression schemes (see Fig.~\ref{fig:ukxfel_slice}).

Electron bunches from the symmetric C-chicane compression scheme have the largest horizontal emittance, which is a result of a sub-optimal CSR mitigation strategy. The transverse projected emittance of electron bunches from the five-dipole chicane compression scheme, and weaker arc-2 compression scheme, is well preserved due to effective mitigation of CSR effects and moderate peak currents. While the arc bunch compressor provides effective CSR mitigation for the lower peak current bunch, it is less effective for the higher peak current bunch. However, the emittance of these electron bunches is still smaller than that of bunches from the symmetric C-chicane compression scheme. The slice emittances (shown in Fig.~\ref{fig:ukxfel_slice}) are well preserved along the electron bunches from each compression scheme, with the exception of slices near the head and tail of electron bunches from the symmetric C-chicane compression scheme.

Unlike in Sec.~\ref{sec:bc_comp_maxiv}, there is an additional linac section (linac 3) following the final bunch compressor. The short range wakefields generated by the electron bunches in this final linac result in different projected energy spreads between the compression cases. For example, the arc compression case results in a projected energy spread that is $83\,\si{\percent}$ larger than the symmetric C-chicane or five-dipole chicane scheme for the $75\,\si{\pico\coulomb}$ case.

The longitudinal phase space distributions of electron bunches at the end of linac 3 are shown in Fig.~\ref{fig:ukxfel_ps} for the arc, symmetric C-chicane and five-dipole chicane compression schemes. The slice properties (current, energy spread and emittance) of electron bunches from each compression scheme are shown in Fig.~\ref{fig:ukxfel_slice}. As the cooperation length is small ($L_c \approx 0.008\,\si{\femto\second}$), the slice properties were calculated for longitudinal slices with a duration of $10 \times L_c$ in order to reduce the statistical errors. 

The projected and slice bunch properties for each compression scheme follow a similar trend to those observed for MAX IV (Sec.~\ref{sec:bc_comp_maxiv}): the projected and slice emittances of electron bunches from the symmetric C-chicane compression schemes are larger than those from the other compression schemes due to poor mitigation of CSR-driven emittance growth; the electron bunches from arc compression schemes typically have larger peak current at equivalent FWTM bunch lengths, as a result of the single-spike current profiles.

\subsection{Free Electron Laser Performance}

The FEL performance for each case is evaluated using \texttt{GENESIS} simulations. Parameters are chosen to produce radiation with a wavelength of $\lambda_r = 62\,\si{\pico\meter}$ (critical photon energy $E_{ph} = 20\,\si{\kilo\electronvolt}$) with an electron beam energy of $8\,\si{\giga\electronvolt}$: the (helical) undulator has a period $\lambda_w = 1.6\,\si{\centi\metre}$ and on-axis magnetic field $B_0 = 0.63\,\si{\tesla}$. The FEL lattice consists of $40$ undulator modules with an active length of $4\,\si{\metre}$ each ($160\,\si{\metre}$ total), separated by drift spaces with a quadrupole at the centre to provide transverse focusing.

\begin{figure}[ht]
	\includegraphics[width=0.95\linewidth]{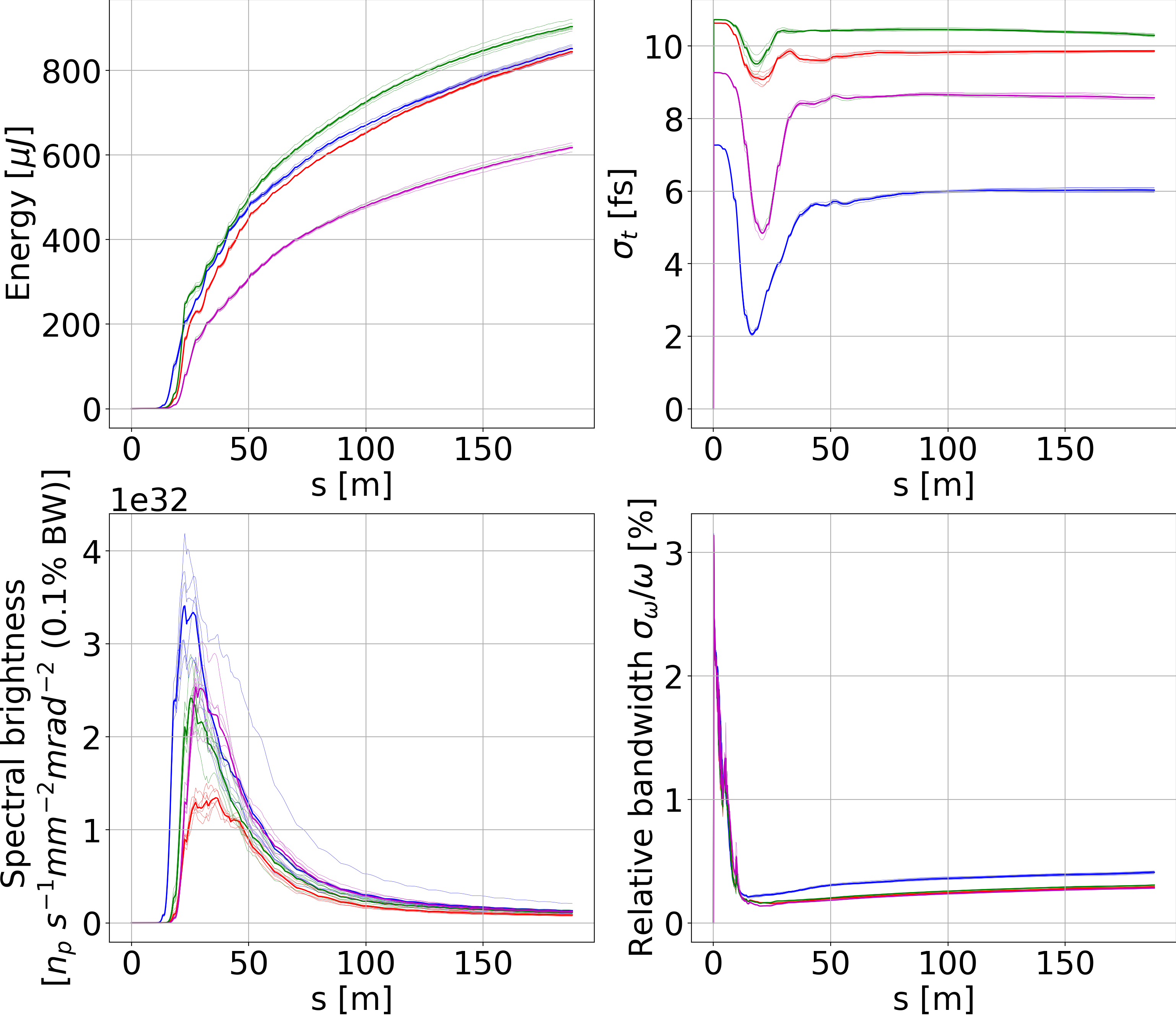}
	\caption{The FEL pulse energy (top left), duration (top right), spectral brightness (bottom left), and relative bandwidth (bottom right) as a function of distance $s$ along the undulator with $75\,\si{\pico\coulomb}$ electron bunches. The plots show \texttt{GENESIS} simulation results for the arc (blue), symmetric C-chicane (red) and five-dipole chicane (green) compression schemes. An additional arc compression case (magenta) is included, which uses a weakly compressed bunch that has the same peak current as the two chicane cases. The thick solid line shows the average of $5$ shots.}
	\label{fig:ukxfel_75pc_genesis_undulator}
\end{figure}

\begin{figure}[ht]
	\includegraphics[width=0.95\linewidth]{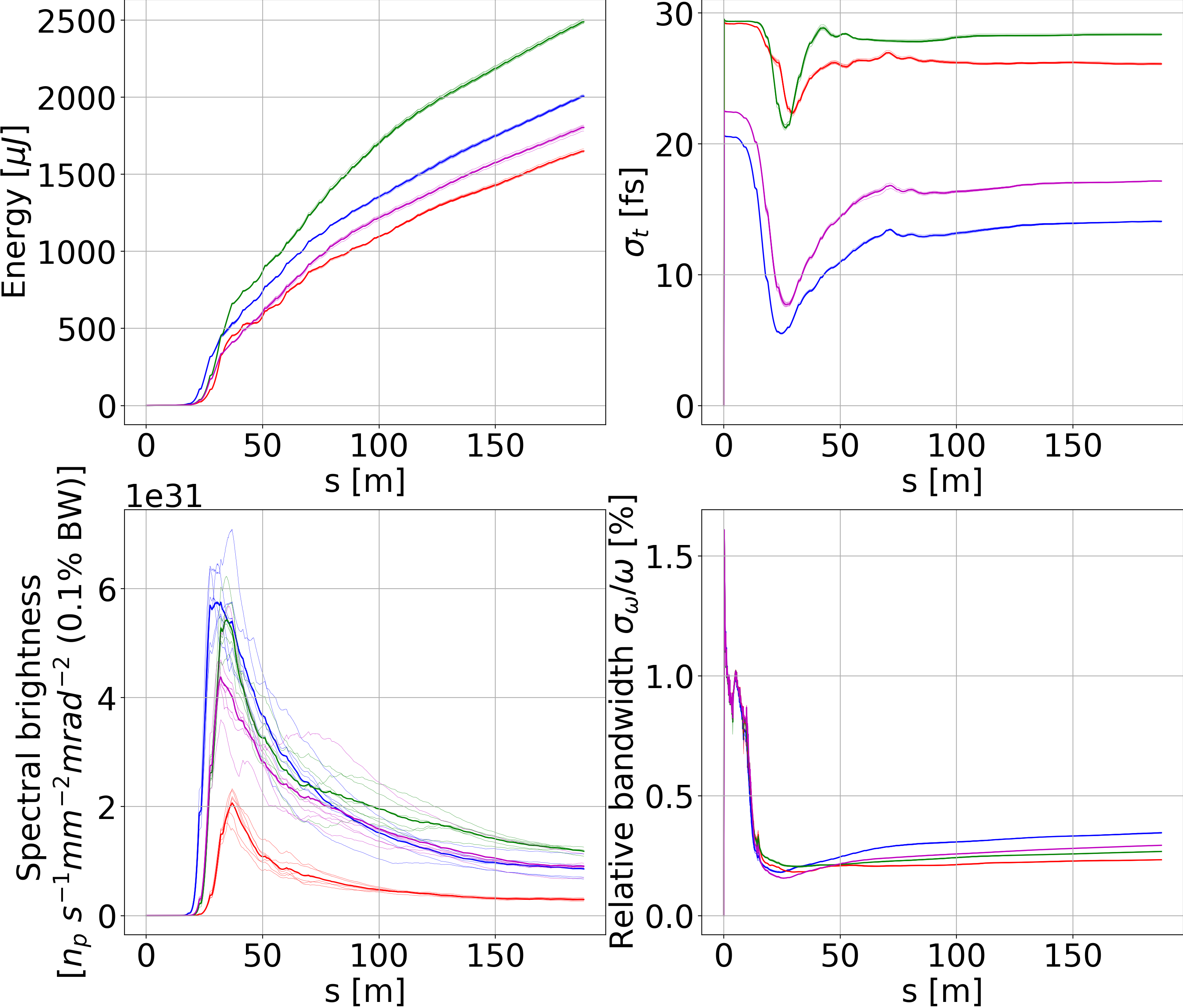}
	\caption{The FEL pulse energy (top left), duration (top right), spectral brightness (bottom left), and relative bandwidth (bottom right) as a function of distance $s$ along the undulator with $300\,\si{\pico\coulomb}$ electron bunches. The plots show \texttt{GENESIS} simulation results for the arc (blue), symmetric C-chicane (red) and five-dipole chicane (green) compression schemes. An additional arc compression case (magenta) is included, which uses a weakly compressed bunch that has the same peak current as the two chicane cases. The thick solid line shows the average of $5$ shots.}
	\label{fig:ukxfel_300pc_genesis_undulator}
\end{figure}

Significant FEL pulse properties (pulse energy, relative bandwidth, pulse duration and spectral brightness) are shown as a function of distance along the undulator in Figs.~\ref{fig:ukxfel_75pc_genesis_undulator} and \ref{fig:ukxfel_300pc_genesis_undulator}.

For a bunch charge of $75\,\si{\pico\coulomb}$, the strong arc (`Arc' case in Tables~\ref{tab:ukxfel_75pc_bunch_properties} and \ref{tab:ukxfel_300pc_bunch_properties}), symmetric C-chicane and five-dipole chicane compression schemes all produce similar pulse energies.
The weak arc compression scheme (arc-2 in Table~\ref{tab:ukxfel_75pc_bunch_properties}) produces significantly lower pulse energies because of the lower average current of the electron bunch. The five-dipole chicane produces the largest pulse energies for both  $75\,\si{\pico\coulomb}$ and  $300\,\si{\pico\coulomb}$ bunch charges as a result of the small slice emittance along the full length of the bunch.

The strong and weak arc compression cases generate FEL pulses with much shorter duration than the pulses from the chicane compression schemes: this is because of the shape of the longitudinal power profiles, which closely follow the current profiles of the electron bunches.

As was the case for MAX IV (Sec.~\ref{sec:bc_comp_maxiv}), the arc compression scheme produces the largest peak spectral brightness due to the high energy and short duration of the FEL pulses. However, the difference in peak spectral brightness is more apparent with $75\,\si{\pico\coulomb}$ bunches (which have high peak current). With $300\,\si{\pico\coulomb}$ bunches, the differences are less significant.

The relative bandwidths of the FEL pulses generated by electron bunches from the strong arc compression case (blue lines in the bottom right plot in Figs.~\ref{fig:ukxfel_75pc_genesis_undulator} and \ref{fig:ukxfel_300pc_genesis_undulator}) are noticeably larger than the other cases, because of the larger projected energy spread of the electron bunches. The other compression schemes produce FEL pulses with relative bandwidths that are very similar to each other.

\begin{figure}[t]
	%\centering
	\includegraphics[width=0.95\linewidth]{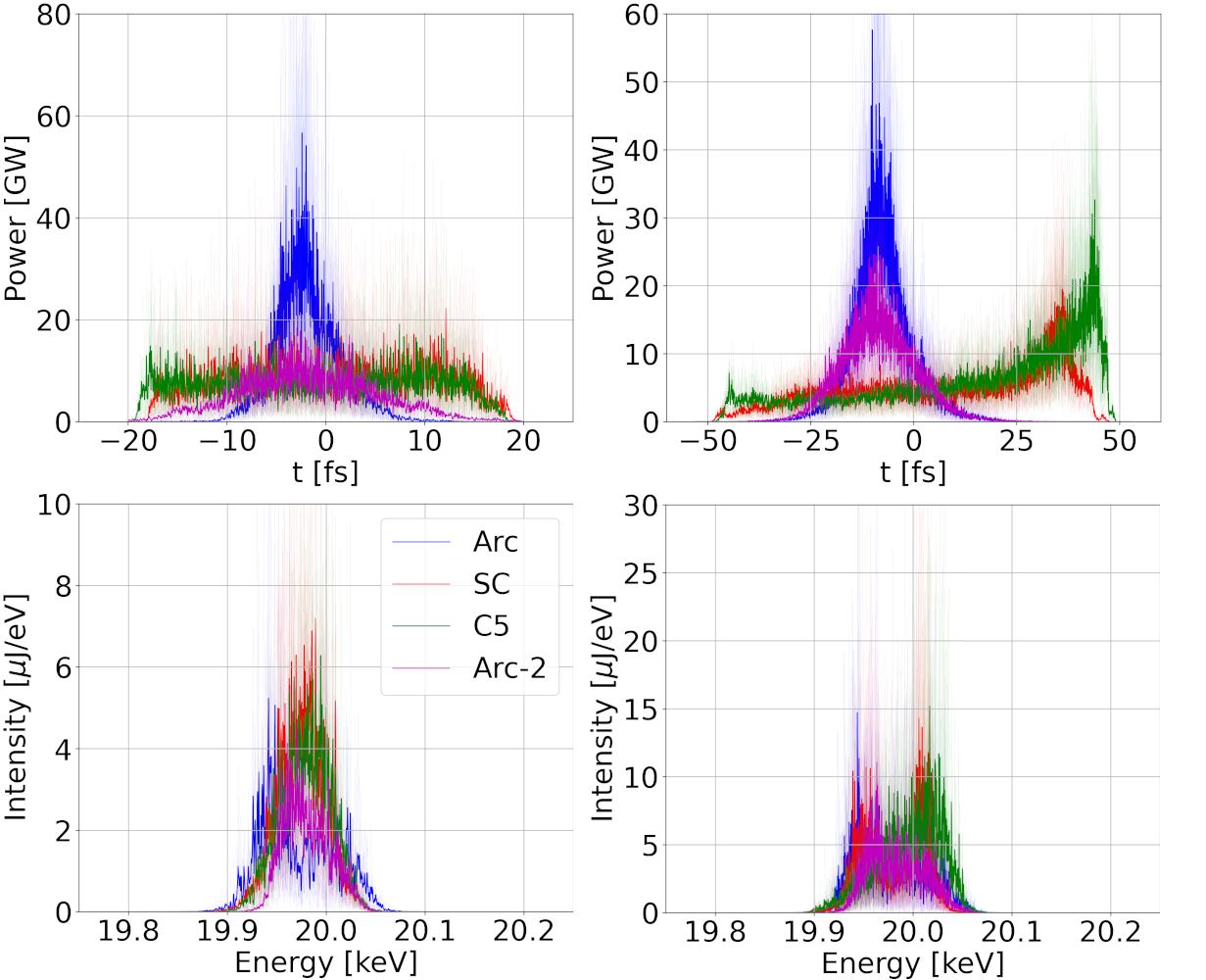}
	\caption{The longitudinal power profiles (top) and energy spectra (bottom) of FEL pulses at peak spectral brightness. Results from simulations using the $75\,\si{\pico\coulomb}$ (left) and $300\,\si{\pico\coulomb}$ (right) electron bunches from the arc (blue), symmetric C-chicane (red), five-dipole chicane (green), and arc-2 (magenta) compression schemes. The thick solid line shows the average of the random shots.}
	\label{fig:ukxfel_genesis_pulse}
\end{figure}

The longitudinal power profiles and energy spectra of FEL pulses at peak spectral brightness are shown in Fig.~\ref{fig:ukxfel_genesis_pulse}. The longitudinal power profile of the FEL pulse produced by the electron bunch from the arc compression scheme is generally more intense over a shorter duration, because of the shape of the current profile of the electron distribution used to generate the FEL pulse. 

Significant properties of the FEL pulse at peak spectral brightness are shown in Tables~\ref{tab:ukxfel_75pc_fel_properties} (for $75\,\si{\pico\coulomb}$ bunches) and \ref{tab:ukxfel_300pc_fel_properties} (for $300\,\si{\pico\coulomb}$ bunches).

\begin{table}[t]
	\caption{FEL pulse properties at peak spectral brightness for each compression scheme using a bunch charge of $75\,\si{\pico\coulomb}$.}
	\begin{ruledtabular}
		\begin{tabular}{l c c c c}
			Pulse properties & Arc & SC & C5 & Arc-2 \\
			\hline
            Spectral brightness & \multirow{2}{2em}{340.6} & \multirow{2}{2em}{134.4} & \multirow{2}{2em}{241.2} & \multirow{2}{2em}{253.5} \\
            {[$10^{30}n_p / \si{\second} / \si{\milli\metre}^2 / \si{\milli\radian}^2 / 0.1 \si{\percent} BW$]} \\
%			\textrm{Spectral brightness [$10^{30}\times$} & & & & \\
%			\textrm{$\si{n_p / \second / \milli\metre^2 / \milli\radian^2 / 0.1 \percent BW}$]} & 340.64 & 134.39 & 241.17 & 253.46 \\
			Energy [\si{\micro\joule}] & 206 & 322 & 272 & 162 \\
%			Energy [\si{\micro\joule}] & 206.35 & 322.40 & 272.37 & 162.11 \\
			Pulse duration $\sigma_t$ [\si{\femto\second}] & 3.25 & 9.68 & 10.15 & 6.68 \\
			Relative bandwidth $\sigma_{\omega}/\omega$ [\si{\percent}] & 0.23 & 0.17 & 0.17 & 0.15 \\
			Peak power [\si{\giga\watt}] & 56.67 & 22.29 & 19.24 & 18.48 \\
			Peak intensity [\si{\micro\joule / \electronvolt}] & 5.24 & 7.20 & 6.29 & 4.50 \\
		\end{tabular}
	\end{ruledtabular}
	\label{tab:ukxfel_75pc_fel_properties}
\end{table}

\begin{table}[t]
	\caption{FEL pulse properties at peak spectral brightness for each compression scheme using a bunch charge of $300\,\si{\pico\coulomb}$.}
	\begin{ruledtabular}
		\begin{tabular}{l c c c c}
			Pulse properties & Arc & SC & C5 & Arc-2 \\
			\hline
            Spectral brightness & \multirow{2}{2em}{57.5} & \multirow{2}{2em}{20.7} & \multirow{2}{2em}{54.3} & \multirow{2}{2em}{43.8} \\
            {[$10^{30}n_p / \si{\second} / \si{\milli\metre}^2 / \si{\milli\radian}^2 / 0.1 \si{\percent} BW$]} \\
%			\textrm{Spectral brightness [$10^{30}\times$} & & & & \\
%			\textrm{$\si{n_p / \second / \milli\metre^2 / \milli\radian^2 / 0.1 \percent BW}$]} & 57.53 & 20.68 & 54.26 & 43.79 \\
			Energy [\si{\micro\joule}] & 391 & 456 & 546 & 329 \\
%			Energy [\si{\micro\joule}] & 391.44 & 455.58 & 545.56 & 329.22 \\
			Pulse duration $\sigma_t$ [\si{\femto\second}] & 6.93 & 25.00 & 26.41 & 9.41 \\
			Relative bandwidth $\sigma_{\omega}/\omega$ [\si{\percent}] & 0.20 & 0.19 & 0.21 & 0.17 \\
			Peak power [\si{\giga\watt}] & 57.67 & 19.54 & 32.66 & 24.63 \\
			Peak intensity [\si{\micro\joule / \electronvolt}] & 14.73 & 14.31 & 15.21 & 11.06 \\
		\end{tabular}
	\end{ruledtabular}
	\label{tab:ukxfel_300pc_fel_properties}
\end{table}

\subsection{Compression Scan}

The compression factor is scanned by varying the RF phase in linac 2 (the linac section between the first and second bunch compressors). The RF voltage is adjusted at the same time as the phase, to maintain a bunch energy of $2.1\,\si{\giga\electronvolt}$ at the end of linac 2. The final bunch lengths approximately cover the range $5\,\si{\femto\second} < \textrm{FWTM} < 90\,\si{\femto\second}$ for the compression schemes with a bunch charge of $75\,\si{\pico\coulomb}$, and a range $20\,\si{\femto\second} < \textrm{FWTM} < 180\,\si{\femto\second}$ in compression schemes with a bunch charge of $300\,\si{\pico\coulomb}$. 

\begin{figure}[t]
	\includegraphics[width=0.493\linewidth]{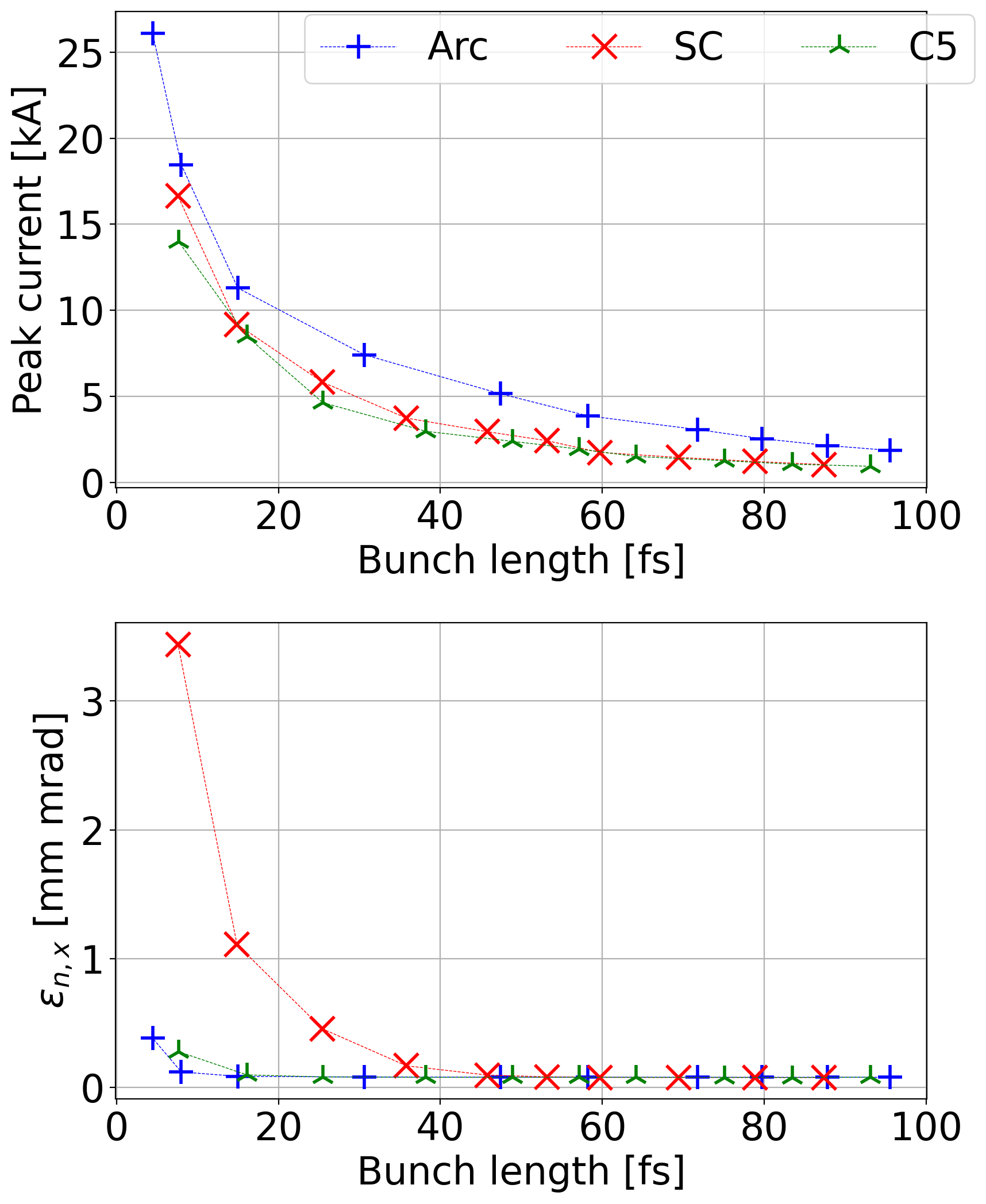}
	\includegraphics[width=0.487\linewidth]{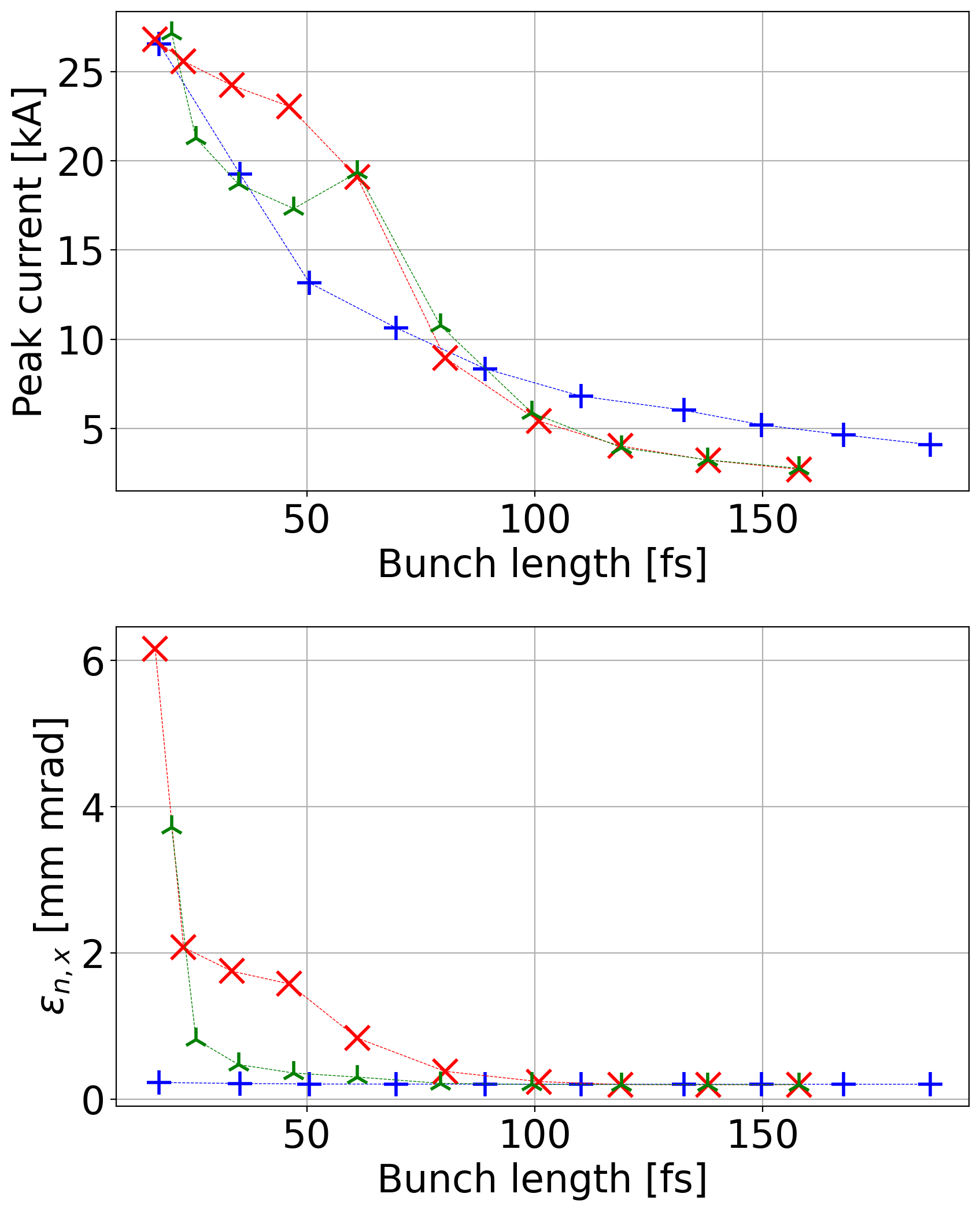}
	\caption{Peak current (top) and horizontal emittance of peak current slice (bottom) as a function of bunch length from compression scans of each compression scheme using bunch charges of $75\,\si{\pico\coulomb}$ (left) and $300\,\si{\pico\coulomb}$ (right). The bunch length is scanned by varying RF phase in linac 2 to control the compression factor in the second bunch compressor.}
	\label{fig:ukxfel_pc_senx}
\end{figure}

The peak current and emittance of the peak current slice are shown as functions of the final bunch length for each compression scheme in Fig.~\ref{fig:ukxfel_pc_senx}.

As was the case for MAX IV (Fig.~\ref{fig:sxl_pc_senx}), with $75\,\si{\pico\coulomb}$ bunch charge the peak currents in the electron bunches from the arc compression scheme are generally higher than the peak currents in either the symmetric C-chicane or five-dipole chicane compression schemes. In the $300\,\si{\pico\coulomb}$ bunch case, the chicane compression schemes have larger currents for bunch lengths below $75\,\si{\femto\second}$, where non-linearities in the longitudinal phase space produce sharp current spikes.

The slice emittance is significantly degraded at stronger compression factors in the symmetric C-chicane bunch compression scheme. In contrast, the emittance of the peak current slice is well preserved in the arc compression scheme, and to a lesser extent in the five-dipole chicane compression scheme, for all compression factors and bunch charges. The preservation of emittance in the arc compression scheme is due to mitigation of CSR-induced emittance growth and the fact that the current profile contains a single spike.

\begin{figure}[t]
	\includegraphics[width=0.493\linewidth]{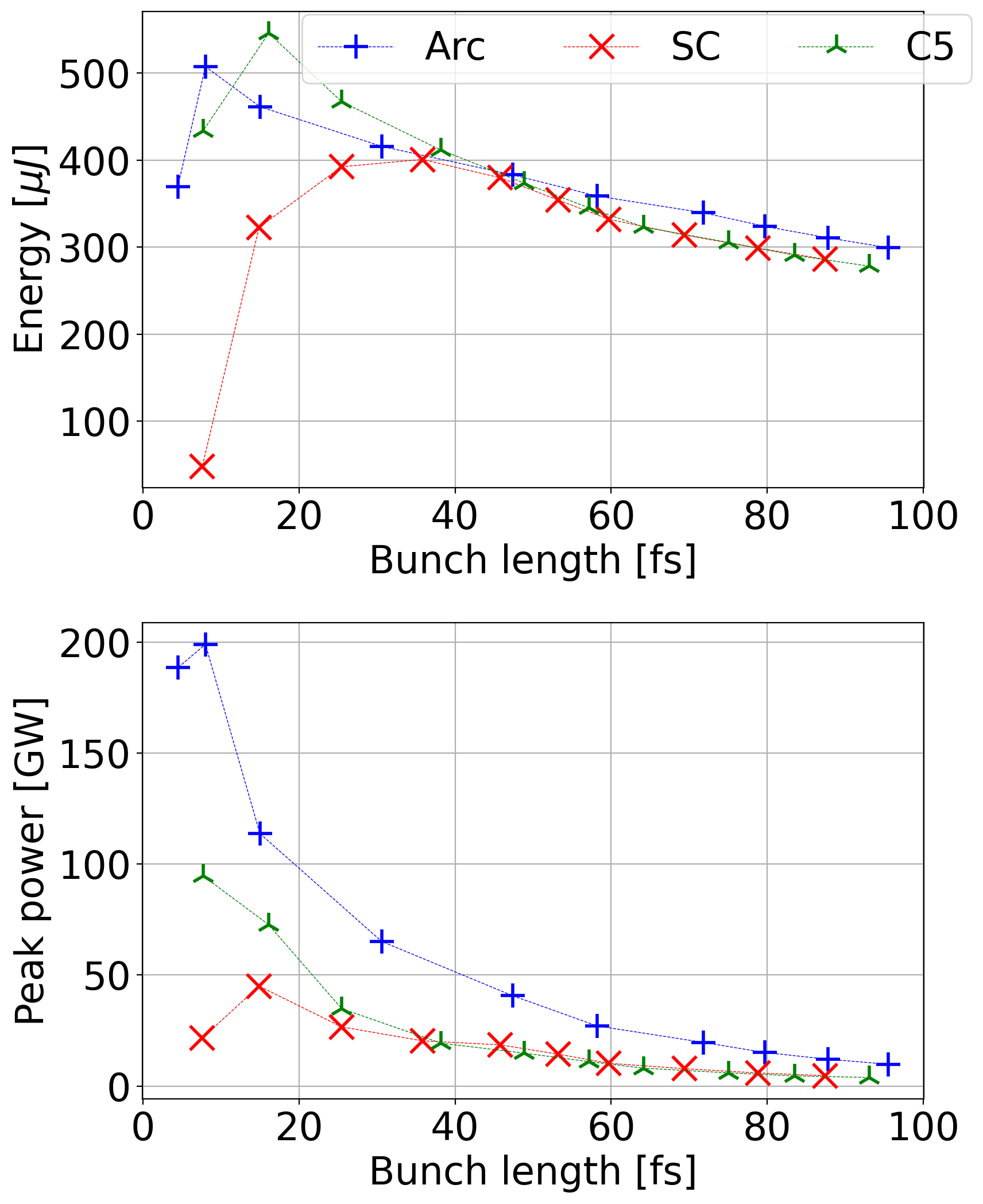}
	\includegraphics[width=0.487\linewidth]{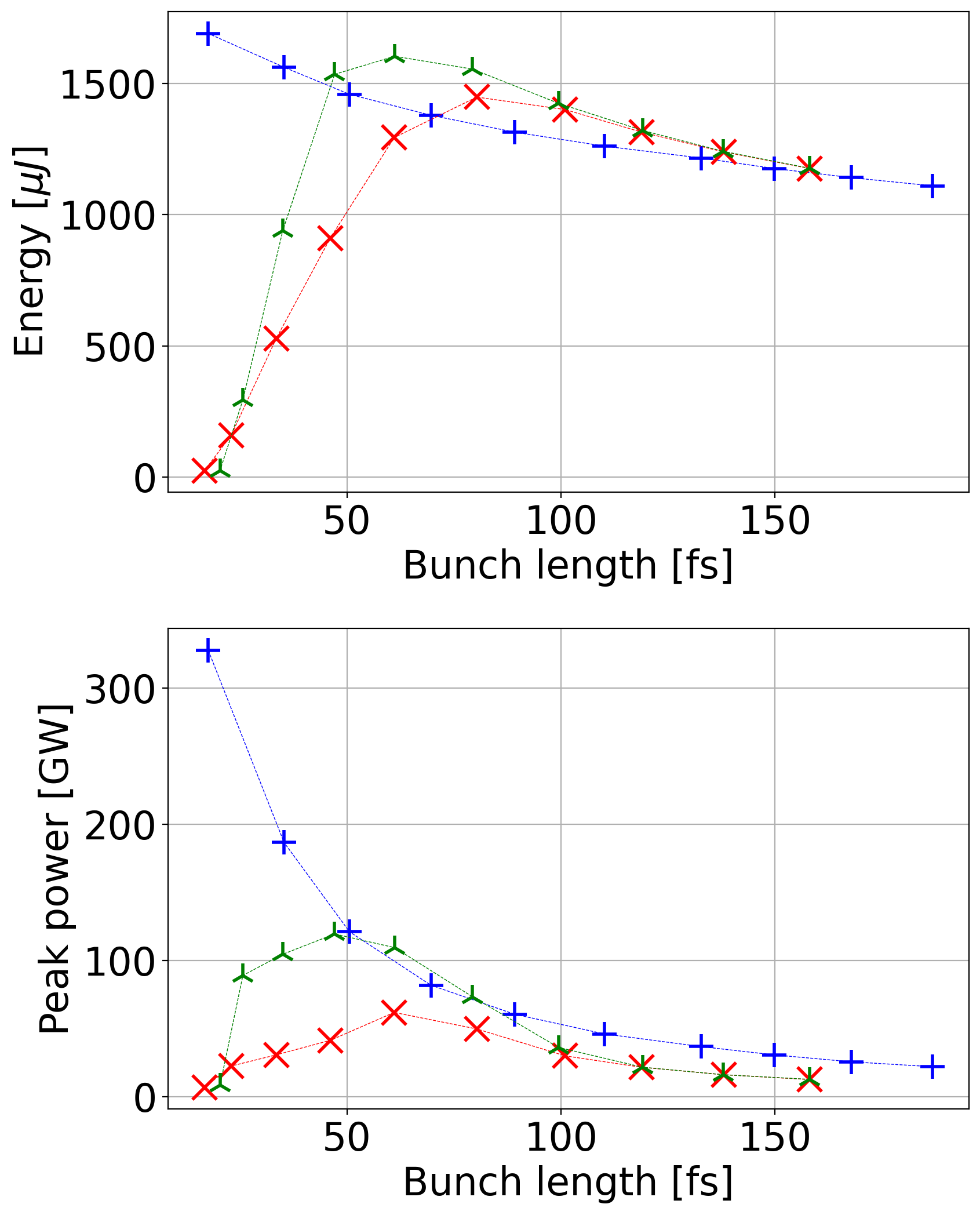}
	\caption{Pulse energy (top) and peak power (bottom) as a function of bunch length from compression scans of each compression scheme using bunch charges of $75\,\si{\pico\coulomb}$ (left) and $300\,\si{\pico\coulomb}$ (right). Pulse energy and peak power are calculated analytically from the slice electron bunch properties. The bunch length is scanned by varying the RF phase in linac 2 to control the compression factor in the second bunch compressor.}
	\label{fig:ukxfel_mingxie}
\end{figure}

The energies and peak powers of FEL pulses generated by electron bunches from each compression scheme are shown in Fig.~\ref{fig:ukxfel_mingxie} as a function of final bunch length. The preservation of slice emittance and high peak currents at short bunch lengths in the arc (and arc-2) compression schemes result in higher peak powers than either of the other compression schemes. However, the differences in peak power and pulse energy between the different bunch compression schemes become insignificant at longer bunch lengths.

\subsection{Charge Sensitivity}

The sensitivity of electron bunch properties to variations in charge were investigated for the different compression schemes using a nominal bunch charges of $75\,\si{\pico\coulomb}$ and $300\,\si{\pico\coulomb}$. The initial bunch charge at the start of linac 1 is varied from $67.5\,\si{\pico\coulomb}$ to $82.5\,\si{\pico\coulomb}$ for the $75\,\si{\pico\coulomb}$ case, and from $270\,\si{\pico\coulomb}$ to $330\,\si{\pico\coulomb}$ for the $300\,\si{\pico\coulomb}$ case. The results are summarised in Tables~\ref{tab:ukxfel_75pc_charge_jitter} ($75\,\si{\pico\coulomb}$) and \ref{tab:ukxfel_300pc_charge_jitter} ($300\,\si{\pico\coulomb}$). 

As in MAX IV (Tables~\ref{tab:sxl_100pc_charge_jitter} and \ref{tab:sxl_10pc_charge_jitter}), the longitudinal bunch properties (bunch length and peak current) in the arc schemes are significantly more sensitive to changes in bunch charge than in either of the chicane compression schemes. The stronger sensitivity is due to a combination of the larger peak current of the electron bunch with nominal charge and the compression stability effect described in Sec.~\ref{sec:bc_comp_maxiv}.
The larger variation in peak current with change in bunch charge results in a greater sensitivity of the projected emittance and energy spread to changes in bunch charge, particularly for the $75\,\si{\pico\coulomb}$ case.

\begin{table}[t]
	\renewcommand{\arraystretch}{1.4}
	%\centering
	\caption{Relative sensitivity of electron bunch properties to variations in bunch charge for each bunch compression scheme with a nominal bunch charge of $75\,\si{\pico\coulomb}$. The sensitivity values are the gradients of a linear fit.}
	\begin{ruledtabular}
		\begin{tabular}{l c c c}
			Charge sensitivity & Arc & SC & C5 \\
			\hline
			${\Delta \textrm{FWTM} / \textrm{FWTM}_{\mathrm{nom}}}/{\Delta Q}$ [$\si{\percent/\pico\coulomb}$] & -4.67 & 1.27 & 0.75 \\
			${\Delta I_{\mathrm{peak}} / I_{\mathrm{peak},\mathrm{nom}}}/{\Delta Q}$ [$\si{\percent/\pico\coulomb}$] & 7.31 & -0.72 & 0.45 \\
			${\Delta \varepsilon_{n,x} / \varepsilon_{n,x,\mathrm{nom}}}/{\Delta Q}$ [$\si{\percent/\pico\coulomb}$] & 5.56 & -0.08 & 0.08 \\
			${\Delta \sigma_{\delta} / \sigma_{\delta,\mathrm{nom}}}/{\Delta Q}$ [$\si{\percent/\pico\coulomb}$] & 1.22 & -0.18 & -0.05 \\
		\end{tabular}
	\end{ruledtabular}
	\label{tab:ukxfel_75pc_charge_jitter}
\end{table}

\begin{table}[t]
	\renewcommand{\arraystretch}{1.4}
	%\centering
	\caption{Relative sensitivity of electron bunch properties to variations in bunch charge for each bunch compression scheme with a nominal bunch charge of $300\,\si{\pico\coulomb}$. The sensitivity values are the gradients of a linear fit.}
	\begin{ruledtabular}
		\begin{tabular}{l c c c}
			Charge sensitivity & Arc & SC & C5 \\
			\hline
			${\Delta \textrm{FWTM} / \textrm{FWTM}_{\mathrm{nom}}}/{\Delta Q}$ [$\si{\percent/\pico\coulomb}$] & -1.18 & 0.37 & 0.37 \\
			${\Delta I_{\mathrm{peak}} / I_{\mathrm{peak},\mathrm{nom}}}/{\Delta Q}$ [$\si{\percent/\pico\coulomb}$] & 7.69 & 0.38 & 0.35 \\
			${\Delta \varepsilon_{n,x} / \varepsilon_{n,x,\mathrm{nom}}}/{\Delta Q}$ [$\si{\percent/\pico\coulomb}$] & 0.75 & 0.35 & 0.09 \\
			${\Delta \sigma_{\delta} / \sigma_{\delta,\mathrm{nom}}}/{\Delta Q}$ [$\si{\percent/\pico\coulomb}$] & 0.12 & -0.08 & -0.08 \\
		\end{tabular}
	\end{ruledtabular}
	\label{tab:ukxfel_300pc_charge_jitter}
\end{table}

\subsection{Microbunching Instability}
\label{sec:mbi}

Microbunching instability (MBI) is of particular concern in high-power X-ray FELs, because the strong emission of CSR from electron bunches with high peak current in the bunch compressor dipoles can significantly increase the energy spread and transverse emittance of the bunches.  The effects can be mitigated by appropriate design of bunch compressor geometry and optics. In this section, we compare MBI in the arc, symmetric C-chicane and five-dipole chicane compression schemes, specifically for the $300\,\si{\pico\coulomb}$ operating mode proposed for UK-XFEL.

An important feature of CSR is the amplification of density modulations that can occur randomly in a bunch entering a bunch compressor.  A useful way to characterise CSR effects in a particular case is to study the amplification factor (ratio of final to initial modulation amplitude) for density modulations of different wavelengths.
To estimate the amplification factor for the bunch compression schemes considered here, simulations were performed in \texttt{ELEGANT} for each compression scheme using electron bunches with `artificial' (and controlled) density modulations applied to the initial distribution. The density modulations had wavelengths of: $10\,\si{\micro\metre}$, $25\,\si{\micro\metre}$, $50\,\si{\micro\metre}$ and $100\,\si{\micro\metre}$. In each of these cases, the relative amplitude of the density modulation was $0.1\,\si{\percent}$.

\begin{figure}[t]
	\includegraphics[width=0.98\linewidth]{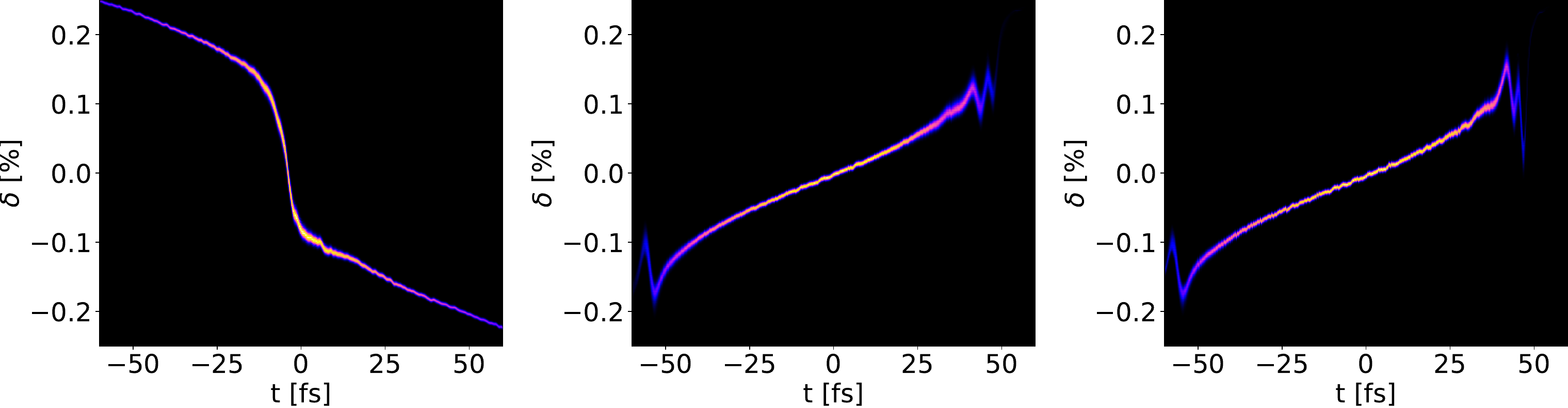}
	\includegraphics[width=0.98\linewidth]{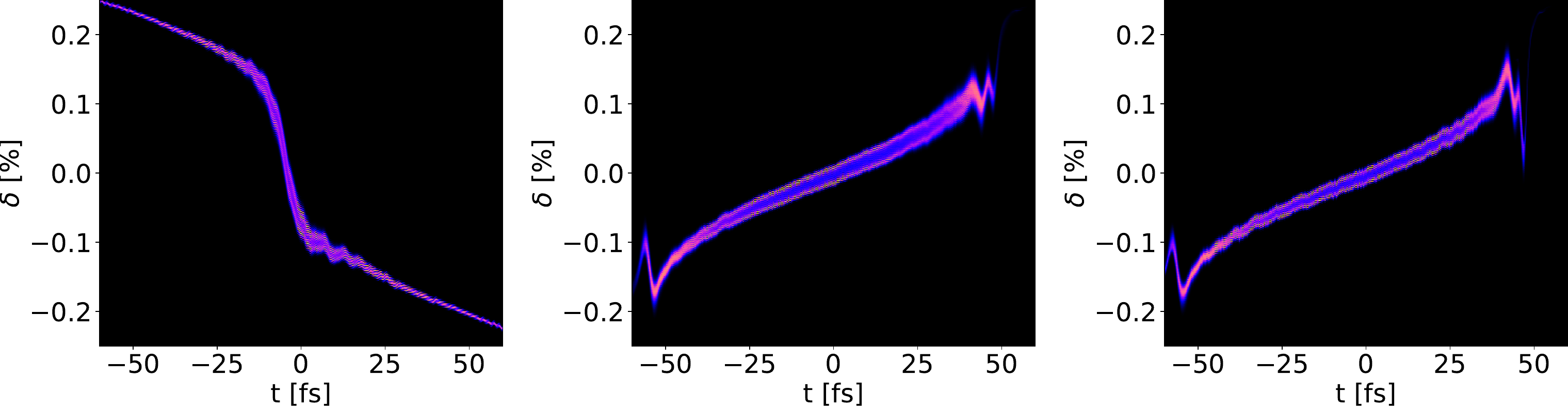}
	\includegraphics[width=0.98\linewidth]{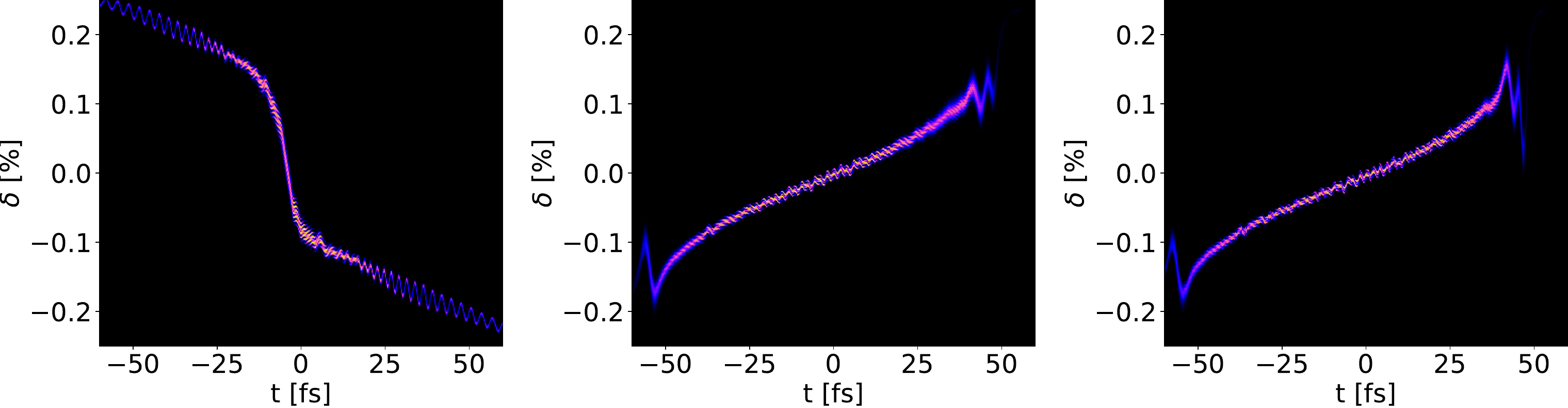}
	\includegraphics[width=0.98\linewidth]{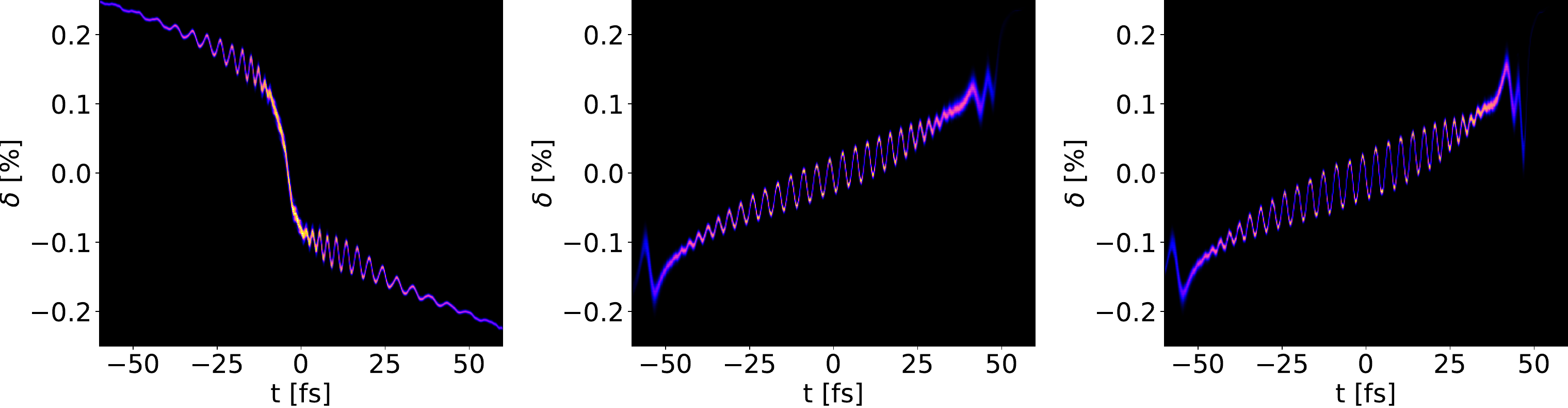}
	\caption{Longitudinal phase space distributions of electron bunches from arc (left), symmetric C-chicane (middle), and five-dipole chicane (right) compression schemes for a bunch charge of $300\,\si{\pico\coulomb}$. The simulated electron bunches had initial density modulations with wavelengths of $10\,\si{\micro\metre}$ (first/top row), $25\,\si{\micro\metre}$ (second row), $50\,\si{\micro\metre}$ (third row), and $100\,\si{\micro\metre}$ (fourth/bottom row). Figure~\ref{fig:ukxfel_ps} shows the case with no initial density modulations.}
	\label{fig:ukxfel_300pc_ps_mbi}
\end{figure}

Figure~\ref{fig:ukxfel_300pc_ps_mbi} shows the longitudinal phase space of electron bunches at the exit of linac 3 (entrance of the FEL) for each compression scheme. The large energy modulations present in the longitudinal phase space distributions result from the short range wakefields in linac 3, which convert the density modulations to energy modulations \cite{huangEffectsLinacWakefield2003}. The energy modulations are larger for the cases with an initial density modulation wavelength of $100\,\si{\micro\metre}$. The longitudinal phase space distributions of electron bunches from the chicane compression schemes have energy and density modulations along the core of the bunch. For the arc compression scheme, the energy and density modulations are mainly present towards the head and tail.

\begin{figure}[t]
	\includegraphics[width=0.98\linewidth]{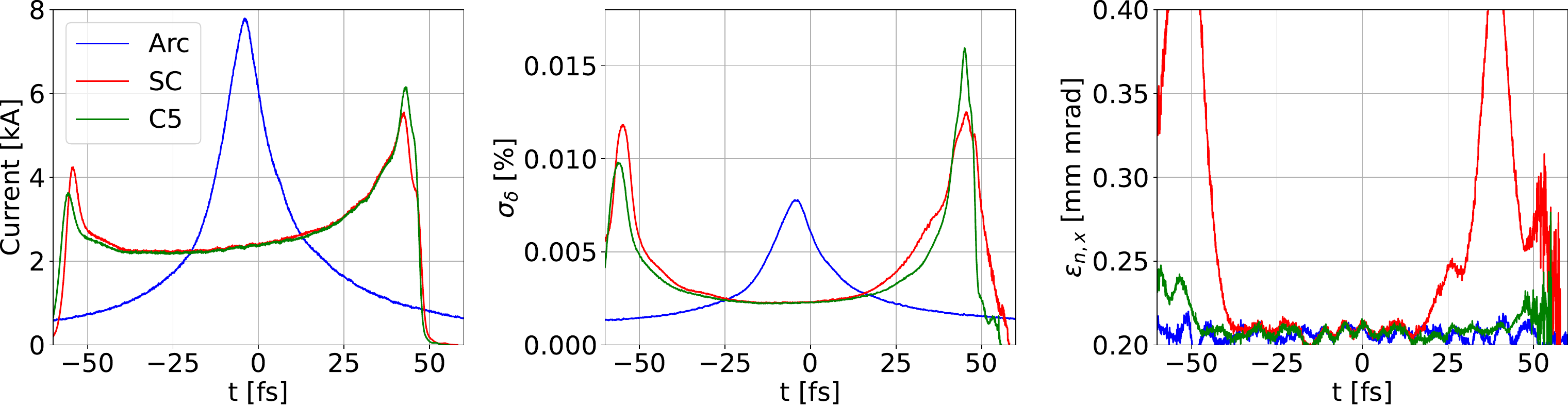}
	\includegraphics[width=0.98\linewidth]{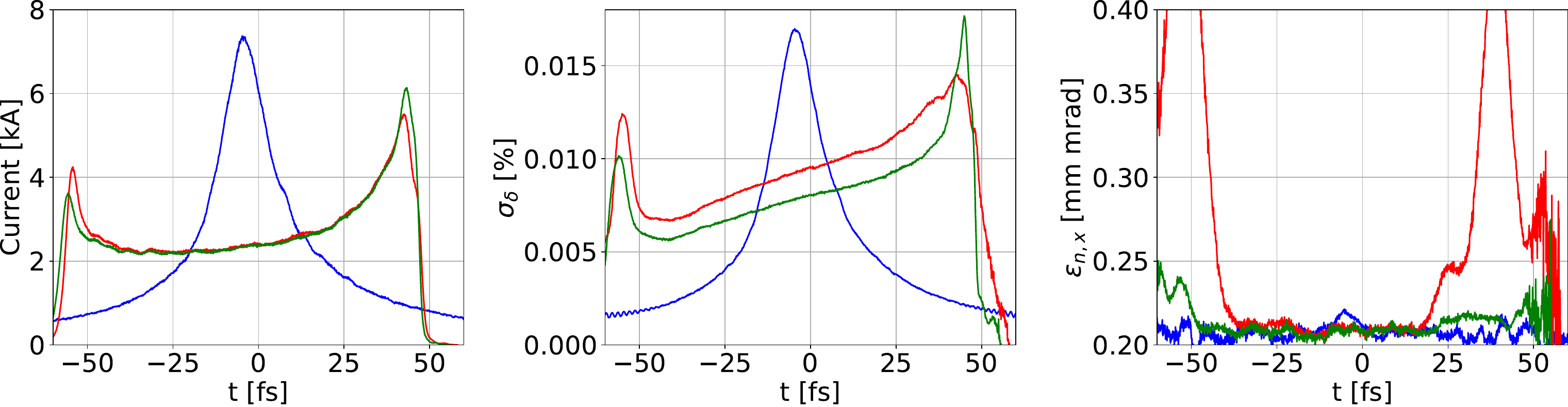}
	\includegraphics[width=0.98\linewidth]{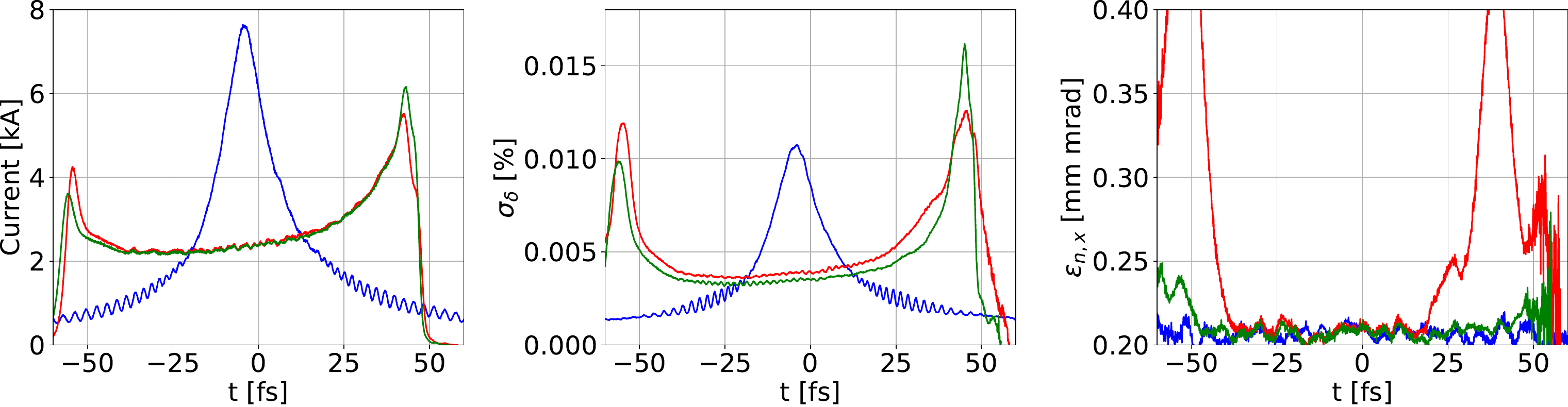}
	\includegraphics[width=0.98\linewidth]{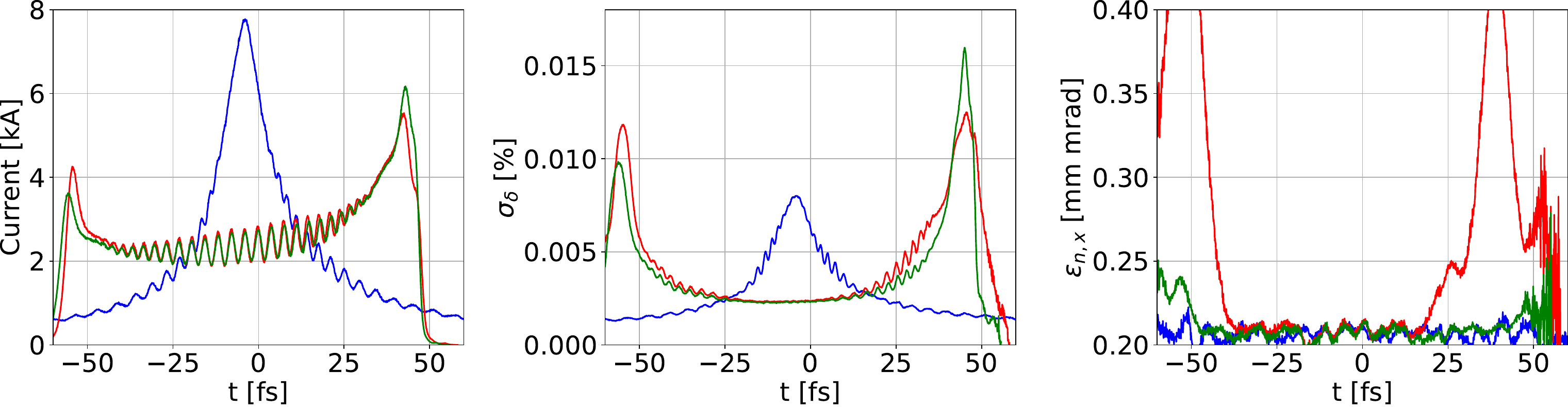}
	\caption{Current (left), slice energy spread (middle) and horizontal slice emittance (right) of electron bunches from arc (blue), symmetric C-chicane (red), and five-dipole chicane (green) compression schemes for a bunch charge of $300\,\si{\pico\coulomb}$. The simulated electron bunches had initial density modulations with wavelengths of $10\,\si{\micro\metre}$ (first/top row), $25\,\si{\micro\metre}$ (second row), $50\,\si{\micro\metre}$ (third row), and $100\,\si{\micro\metre}$ (fourth/bottom row). Figure~\ref{fig:ukxfel_slice} shows the case with no initial density modulations.}
	\label{fig:ukxfel_300pc_slice_mbi}
\end{figure}

From the current profiles of the electron distributions from each compression scheme (shown in the left column of Fig.~\ref{fig:ukxfel_300pc_slice_mbi}) it can be seen that density modulations are larger for the cases that had an initial density modulation wavelength of $100\,\si{\micro\metre}$. Density modulations are present along the core of an electron bunch from the five-dipole chicane compression scheme, while density modulations are mainly present towards the head and tail of a bunch from the arc compression scheme. The absence of density modulations in the core of a bunch from the arc compression scheme is due to the larger slice energy spread (with the contribution of the chirp to the slice energy spread) in this region: this provides Landau damping, which suppresses microbunching \cite{huangSuppressionMicrobunchingInstability2004}.

As the current at the head and tail of a bunch in the arc compression scheme is relatively low, these regions typically make a smaller contribution to the lasing in an FEL than the core of the electron bunch. Therefore, the energy and density modulations at the head and tail of electron bunches from the arc compression scheme will not significantly impact the FEL performance. However, the density and energy modulations that are present in the core of an electron bunch from the five-dipole chicane compression scheme, which makes a significant contribution to the lasing in an FEL, will have an adverse impact on FEL performance.

\section{Discussion}

The results presented in this paper characterise the impact on bunch properties of beam dynamics effects in different types of bunch compressor in linac driven FELs. Arc, symmetric C-chicane and five-dipole chicane compression schemes were investigated for two FEL regimes: a soft X-ray FEL driven by a normal-conducting S-band linac, and a hard X-ray FEL driven by a super-conducting linac.

The symmetric C-chicane compression schemes typically produce electron bunches with larger horizontal emittances than either the five-dipole chicane or arc compression schemes; this was demonstrated for a range of bunch charges, bunch lengths, and compression factors. At strong compression factors, the degradation is even more significant. The five-dipole chicane and arc bunch compressors are both more effective at mitigating CSR-induced emittance growth than the symmetric C-chicane. The vertical emittances, and horizontal emittances in cases where CSR effects are negligible, may be larger for electron bunches from the arc compression scheme due to chromatic effects. In principle, chromatic effects can be mitigated further through the addition of linear and non-linear optical elements \cite{lindstromDesignGeneralApochromatic2016,bjorklundsvenssonThirdOrderDoubleAchromatBunch2019}.

The smaller transverse emittance from the five-dipole chicane and arc compression schemes typically generate FEL pulses that have higher pulse energies and higher peak spectral brightness than those generated from the symmetric C-chicane. Additionally, the relative bandwidth is larger when driven by a linac with an arc-like compression scheme, particularly when the final bunch compressor is followed by an additional linac.  However, energy spread can be reduced using corrugated wakefield dechirpers \cite{baneCorrugatedPipeBeam2012,zhangElectronBeamEnergy2015}, and dielectric lined waveguides can be used to remove a positive chirp from arc-like compression schemes \cite{paceySimulationStudiesDielectric2018}.

The longitudinal beam dynamics in an arc-like compression scheme differs from a chicane-like compression scheme because of the method used to linearise the longitudinal phase space, short range wakefields, and CSR. These differences result in electron bunches with different current profiles. The arc compression schemes typically result in Gaussian-like current profiles, and the chicane schemes result in more uniform current profiles. Current spikes are often present at the head and tail of bunches from a chicane-like compression scheme. The different current profiles are suitable for different FEL schemes. For example, an ultra-short single spike in the current profile of an electron bunch, which is readily achieved in an arc compression scheme, can generate a single coherent spike in the longitudinal profile of an FEL pulse. A single-spike FEL pulse is useful for attosecond FEL schemes. However, the sharp spike in the current profile of an electron bunch from an arc compression scheme may be less suitable for HB-SASE, which requires the longitudinal slices to reach saturation at the same time. It is worth mentioning that the observation that arc-like compression is more suitable for attosecond schemes may only apply to the compression schemes studied in this paper, as other techniques can also be used to generate short coherent FEL pulses, such as emittance spoiling \cite{emmaFemtosecondSubfemtosecondXRay2004}. Further work may be needed to investigate the different bunch compression schemes used for different FEL schemes, such as HB-SASE or self-seeding.

Studies have been conducted to investigate the impact of MBI on a proposed design for a UK-XFEL. Simulations of the compression schemes were performed using initial electron bunches with small density modulations at various modulation wavelengths. After compression, the bunches from the arc compression scheme did not have density or energy modulations in the core of the bunch (i.e.~in the region of the current spike), because the larger slice energy spread in this region leads to Landau damping that suppresses microbunching gain \cite{huangSuppressionMicrobunchingInstability2004, saldinLongitudinalSpaceChargeDriven2004}. In contrast, electron bunches from the five-dipole chicane compression scheme had density and energy modulations along the core of the bunch. While further work is needed to understand the differences in MBI between the different compression schemes, these initial results indicate that an arc compression scheme will be less susceptible to MBI than a chicane compression scheme. Simulations of the compression schemes proposed for UK-XFEL did not lead to MBI when the initial electron bunches did not have initial density or energy modulations.

The sensitivity of significant electron bunch properties to variations in bunch charge were investigated for each compression scheme. The results showed that the longitudinal properties (bunch length and peak current) were more sensitive to variations in charge for the arc compression schemes than in the chicane compression schemes. The stronger sensitivities of bunch length and peak current to variations in charge for the arc-like compression schemes was related to the larger peak current at the nominal bunch charge and bunch length for the arc compression scheme compared to the chicane compression schemes. The sensitivity is enhanced by the fact that short range wakefields and CSR increase the energy spread in electron bunches in the arc compression scheme, which increases the total compression factor in a two-stage compression scheme. Two-stage chicane-like compression scheme will be less strongly affected by variations in bunch charge.

\section{Summary}

A symmetric C-chicane bunch compression scheme is sub-optimal for delivering high peak current, low emittance electron bunches to a free electron laser, as CSR has significant effects on the transverse electron distribution. Compression schemes using either five-dipole chicanes or arc compressors (with CSR mitigating optics) are more suited to delivering low emittance bunches to a free electron laser, as they are capable of mitigating CSR effects and preserving the transverse emittance. While both five-dipole chicanes and arc bunch compressors are able to preserve the transverse emittance, the differences in the longitudinal beam dynamics between an arc-like compression scheme and a chicane-like compression scheme result in different longitudinal phase space distributions.  Arc-like compression schemes use bunches with opposite sign chirp to those in chicane compression schemes, and produce current profiles that have a single spike.  The current profile in bunches produced by chicane compression schemes are more uniform in the core, but have current spikes at the head and tail of the bunch. The differences in the current profiles mean that a chicane-like compression scheme will be more appropriate for FEL schemes that require small energy spread and uniform current profile (such as HB-SASE or self-seeding) while an arc-like compression scheme will be more suitable for attosecond schemes. Because of this difference between the potential applications of chicane-like compression schemes and arc-like compression schemes, we recommend that facilities aiming to drive multiple FEL beamlines from a single linac, such as UK-XFEL, be designed and constructed with the flexibility to select between arc-like compression and chicane-like compression on a bunch-by-bunch basis. One possible system for achieving this has been proposed in this paper.

\section{Acknowledgements}
This work was supported by the Science and Technology Facilities Council, U.K., through a grant to the Cockcroft Institute.

\bibliography{References}% Produces the bibliography via BibTeX.

\end{document}